%% file: BayesHerwigTune-arxiv.tex
\documentclass[zpreprint,zbstnp,final]{zeus_paper}
\usepackage{placeins}
\usepackage[titletoc,title]{appendix}
\usepackage{authblk}
\usepackage{hyperref}
\usepackage{calc}
\usepackage[utf8]{inputenc}
\usepackage{graphicx}
\usepackage{titlesec}
\usepackage{xspace}
\usepackage{setspace}
\usepackage{subcaption}
\usepackage{amsthm}
\usepackage{mcite}
\usepackage{url}

\newcommand*{\ditto}{---\texttt{"}---\xspace}
\graphicspath{{./Figures/}{./Figures/tune_plots/}{./Figures/error_prop/}{./Figures/int_plots/}{./Figures/weight_plots/}{./Figures/correlation_plots/}{./Figures/tuned_obs/}}

\title{A Bayesian tune of the Herwig Monte Carlo event generator}
\author[a]{Salvatore La Cagnina\thanks{salvatore.lacagnina@tu-dortmund.de}}
\author[a]{Kevin Kr\"oninger\thanks{kevin.kroeninger@cern.ch}}
\author[b]{\\Stefan Kluth\thanks{stefan.kluth@mpp.mpg.de}}
\author[b]{Andrii Verbytskyi\thanks{andrii.verbytskyi@mpp.mpg.de}}
\affil[a]{Technische Universit\"at Dortmund,  Fakult\"at Physik, \protect\\ Otto-Hahn-Stra{\ss}e 4, D-44227 Dortmund, Germany}
\affil[b]{Max-Planck-Institut f\"ur Physik, \protect\\ F\"ohringer Ring 6, D-80805 Munich, Germany}

\newcommand{\epjcbreak}[1]{}
\newcommand{\draftbreak}[1]{}

\newcommand{\epjconly}[1]{}
\newcommand{\draftonly}[1]{}

\include{BayesHerwigTune-tab}
\include{BayesHerwigTune-fig}

\include{BayesHerwigTune-def}

\begin{document}
\abstract{\input{BayesHerwigTune-abs}}
\makezeustitle 
\newpage
\clearpage
\pagenumbering{arabic}
\pagestyle{plain}
\include{BayesHerwigTune-txt}
{\bibliographystyle{BayesHerwigTune}{\raggedright\bibliography{BayesHerwigTune.bib}}}\vfill\eject
\clearpage
\end{document}

%% file: BayesHerwigTune-tab.tex
\newcommand{\TABparamsher}{
\begin{table}[hbp]
\centering
\begin{tabular}{| c | c | c |}\hline
    Parameter & Range & Default \\ \hline\hline
AlphaQCD      &     $[0.1000, 0.1417]$  &   0.1181  \\ 
IRcutoff (GeV) &     $[0.5004, 1.5012]$    &   1.0080   \\ 
$m_{g}$ (GeV) &     $[0.7445, 1.1400]$    &   0.9500    \\ 
$m_{s}$ (GeV) &     $[0.4734, 0.5000] m_{g}$   &   0.4500    \\ 
ClMax (GeV) &     $[1.9334, 5.8000]$    &   3.8667  \\ 
ClPow       &     $[0.8295, 2.4885]$    &   1.6590   \\ 
ClSmr       &     $[0.1719, 0.8593]$    &   0.3437  \\ 
PSplit      &     $[0.3450, 1.0348]$    &   0.6899  \\ \hline
\end{tabular}
\caption{Parameters for the \texttt{Herwig7-H7} tune, their ranges and default values. The quantities without units are dimensionless.}
\label{table:parameterH7}
\end{table}
}

\newcommand{\TABlonglistone}[1]{
\begin{table}[!h]
\centering
\resizebox{0.9\textwidth}{!}{  
\begin{tabular}{| c | c | c | c |}\hline
 \texttt{Rivet} analysis and bin code     &    Description       &  \multicolumn{2}{c|} {Weight scheme } \\ 
                 &                     & $w_1$ &  $w_2$ \\ \hline\hline

\multicolumn{4}{|c|}{ALEPH\_1996\_S3486095~\cite{ALEPH:1996oqp}} \\\hline
  d01-x01-y01 & Sphericity, $S$ (charged) & 1 & 5 \\
  d02-x01-y01 & Aplanarity, $A$ (charged) & 2 & 10 \\
  d03-x01-y01 & 1-Thrust, $1-T$ (charged) & 1 & 5 \\
  d04-x01-y01 & Thrust minor, $m$ (charged) & 2 & 10 \\
  d07-x01-y01 & $C$ parameter (charged) & 1 & 5 \\
  d08-x01-y01 & Oblateness, $M - m$ (charged) & 1 & 5 \\
  d09-x01-y01 & Scaled momentum, $x_p = |p|/|p_\text{beam}|$ (charged)  & 1 & 5 \\
  d11-x01-y01 & In-plane $p_T$ w.r.t. sphericity axes (charged) & 1 & 5  \\
  d12-x01-y01 & Out-of-plane $p_T$ w.r.t. sphericity axes (charged)  & 1 & 5 \\
  d17-x01-y01 & Log of scaled momentum, $\log(1/x_p)$ (charged) & 1 & 5 \\
  d18-x01-y01 & Charged multiplicity & 2 & 10 \\
  d19-x01-y01 & Mean charged multiplicity & 150 & 750 \\
  d25-x01-y01 & $\pi^\pm$ spectrum & 1 & 1 \\
  d26-x01-y01 & $K^\pm$ spectrum & 1 & 1 \\
  d29-x01-y01 & $\pi^0$ spectrum & 1 & 1 \\
  d30-x01-y01 & $\eta$ spectrum & 1 & 1 \\
  d31-x01-y01 & $\eta'$ spectrum & 1 & 1 \\
  d32-x01-y01 & $K^0$ spectrum & 1 & 1  \\
  d33-x01-y01 & $\Lambda^0$ spectrum & 1 & 1 \\
d34-x01-y01 & $\Xi^-$ spectrum & 1 & 1 \\
  d35-x01-y01 & $\Sigma^\pm(1385)$ spectrum & 1 & 1 \\
  d36-x01-y01 & $\Xi^0(1530)$ spectrum & 1 & 1 \\
  d37-x01-y01 & $\rho$ spectrum & 1 & 1  \\
  d38-x01-y01 & $\omega(782)$ spectrum & 1 & 1 \\
  d39-x01-y01 & $K^{*0}(892)$ spectrum  & 1 & 1 \\
  d40-x01-y01 & $\phi$ spectrum & 1 & 1 \\
  d43-x01-y01 & $K^{*\pm}(892)$ spectrum & 1 & 1 \\\hline
\multicolumn{4}{|c|}{ALEPH\_2001\_S4656318~\cite{ALEPH:2001pfo}}\\\hline 
 d01-x01-y01 & $b$ quark fragmentation  function   $f(x_B^\text{weak})$  & 7 & 35 \\
 d07-x01-y01 & Mean of $b$ quark fragmentation function $f(x_B^\text{weak})$ & 3 & 15  \\\hline
\multicolumn{4}{|c|}{JADE\_OPAL\_2000\_S4300807~\cite{JADE:1999zar}}\\\hline
  d26-x01-y01 & 2-jet Durham diff. rate & 2 & 10 \\
  d26-x01-y02 & 3-jet Durham diff. rate & 2 & 10 \\
  d26-x01-y03 & 4-jet Durham diff. rate & 2 & 10 \\
  d26-x01-y04 & 5-jet Durham diff. rate & 2 & 10 \\\hline

\end{tabular}
}
\caption{#1}
\label{table:longlistone}
\end{table}
}

\newcommand{\TABlonglistthree}[1]{
\begin{table}[!h]
\centering
\resizebox{0.90\textwidth}{!}{  
\begin{tabular}{| c | c | c | c |}\hline
 \texttt{Rivet} analysis and bin code     &    Description       &  \multicolumn{2}{c|} {Weight scheme } \\ 
                 &                     & $w_1$ &  $w_2$ \\ \hline\hline
\multicolumn{4}{|c|}{PDG\_HADRON\_MULTIPLICITIES~\cite{ParticleDataGroup:2008zun}} \\\hline  
 d40-x01-y02 & \ditto $\Sigma^0$ & 10 & 0 \\
 d41-x01-y01 & \ditto $\Sigma^-$ & 10 & 0 \\
 d42-x01-y01 & \ditto $\Sigma^+$ & 10 & 0 \\
 d43-x01-y01 & \ditto $\Sigma^\pm$ & 10 & 0 \\
 d44-x01-y03 & \ditto $\Xi^-$ & 10 & 0 \\
 d45-x01-y02 & \ditto $\Delta^{++}(1232)$ & 10 & 0 \\
 d46-x01-y03 & \ditto $\Sigma^-(1385)$ & 10 & 0 \\
 d47-x01-y03 & \ditto $\Sigma^+(1385)$ & 10 & 0 \\
 d48-x01-y03 & \ditto $\Sigma^\pm(1385)$ & 10 & 0 \\
 d49-x01-y02 & \ditto $\Xi^0(1530)$ & 10 & 0 \\
 d50-x01-y03 & \ditto $\Omega^-$ & 10 & 0 \\
 d51-x01-y03 & \ditto $\Lambda_c^+$ & 10 & 0 \\
 d52-x01-y01 & \ditto $\Lambda_b^0$ & 10 & 0 \\
 d54-x01-y02 & \ditto $\Lambda(1520)$ & 10 & 0 \\\hline            
\multicolumn{4}{|c|}{DELPHI\_1996\_S3430090~\cite{DELPHI:1996sen}}\\\hline 
d01-x01-y01 & In-plane $p_\perp$ w.r.t.  thrust axes & 1 & 5 \\
  d02-x01-y01 & Out-of-plane $p_\perp$ w.r.t.  thrust axes & 1 & 5 \\
  d03-x01-y01 & In-plane $p_\perp$ w.r.t.  sphericity axes  & 1 & 5 \\
  d04-x01-y01 & Out-of-plane $p_\perp$ w.r.t.  sphericity axes  & 1 & 5 \\
  d07-x01-y01 & Scaled momentum,  $x_p = |p|/|p_\text{beam}|$ & 1 & 5 \\

  d08-x01-y01 & Log of scaled momentum, $\log(1/x_p)$& 1 & 5 \\
  d09-x01-y01 & Mean out-of-plane $p_\perp$ w.r.t. thrust axes vs. $x_p$ & 1 & 5 \\
  d10-x01-y01 & Mean $p_\perp$ vs. $x_p$ & 1 & 5 \\
  d11-x01-y01 & $1-\text{Thrust}$ & 1 & 5 \\
  d12-x01-y01 & Thrust major, $M$ & 1 & 5 \\
  d13-x01-y01 & Thrust minor, $m$ & 2 & 10 \\
  d14-x01-y01 & Oblateness = $M - m$ & 1 & 5 \\
  d15-x01-y01 & Sphericity, $S$ & 1 & 5 \\
  d16-x01-y01 & Aplanarity, $A$ & 2 & 10 \\
  d17-x01-y01 & Planarity, $P$ & 1 & 5 \\
  d18-x01-y01 & $C$ parameter & 1 & 5 \\
  d19-x01-y01 & $D$ parameter & 1 & 5 \\
  d33-x01-y01 & Energy-energy correlation, EEC  & 1 & 5 \\
  d35-x01-y01 & Mean charged multiplicity & 150 & 750 \\\hline

\end{tabular}
}
\caption{#1}
\label{table:longlistthree}
\end{table}
}

\newcommand{\TABlonglisttwonew}[1]{
\begin{table}[!h]
\centering
\resizebox{0.69\textwidth}{!}{  
\begin{tabular}{| c | c | c | c |}\hline
 \texttt{Rivet} analysis and bin code     &    Description       &  \multicolumn{2}{c|} {Weight scheme } \\ 
                 &                     & $w_1$ &  $w_2$ \\ \hline\hline

\multicolumn{4}{|c|}{PDG\_HADRON\_MULTIPLICITIES~\cite{ParticleDataGroup:2008zun}} \\\hline  
 d01-x01-y03 &Multiplicity of $\pi^+$ & 10 & 0 \\
 d02-x01-y03 & \ditto $\pi^0$ & 10 & 0 \\
 d03-x01-y03 & \ditto $K^+$ & 10 & 0 \\
 d04-x01-y03 & \ditto $K^0$ & 10 & 0 \\
 d05-x01-y03 & \ditto $\eta$ & 10 & 0 \\
 d06-x01-y03 & \ditto $\eta'(958)$ & 10 & 0 \\
 d07-x01-y03 & \ditto $D^+$ & 10 & 0 \\
 d08-x01-y03 & \ditto $D^0$ & 10 & 0      \\
 d09-x01-y03 & \ditto $D^+_s$ & 10 & 0 \\
 d10-x01-y01 & \ditto $B^+, B^0_d$ & 10 & 0 \\
 d11-x01-y01 & \ditto $B^+_u$ & 10 & 0 \\
 d12-x01-y01 & \ditto $B^0_s$ & 10 & 0 \\
 d13-x01-y03 & \ditto $f_0(980)$ & 10 & 0 \\
 d14-x01-y01 & \ditto $a_0^+(980)$ & 10 & 0 \\
 d15-x01-y03 & \ditto $\rho^0(770)$ & 10 & 0 \\
 d16-x01-y01 & \ditto $\rho^+(770)$ & 10 & 0 \\
 d17-x01-y02 & \ditto $\omega(782)$ & 10 & 0 \\
 d18-x01-y03 & \ditto $K^{*+}(892)$ & 10 & 0 \\
 d19-x01-y03 & \ditto $K^{*0}(892)$ & 10 & 0 \\
 d20-x01-y03 & \ditto $\phi(1020)$ & 10 & 0 \\
 d21-x01-y03 & \ditto $D^{*+}(2010)$ & 10 & 0 \\
 d23-x01-y02 & \ditto $D^{*+}_s(2112)$ & 10 & 0 \\
 d24-x01-y01 & \ditto $B^*$ & 10 & 0 \\
 d25-x01-y02 & \ditto $J/\psi(1S)$ & 10 & 0 \\
 d26-x01-y01 & \ditto $\psi(2S)$ & 10 & 0 \\
 d27-x01-y01 & \ditto $\Upsilon(1S)$ & 10 & 0 \\
 d28-x01-y01 & \ditto $f_1(1285)$ & 10 & 0 \\
 d29-x01-y01 & \ditto $f_1(1420)$ & 10 & 0 \\
 d30-x01-y01 & \ditto $\chi_{c1}(3510)$ & 10 & 0 \\
 d31-x01-y03 & \ditto $f_2(1270)$ & 10 & 0 \\
 d32-x01-y01 & \ditto $f_2'(1525)$ & 10 & 0 \\
 d34-x01-y02 & \ditto $K_2^{*0}(1430)$ & 10 & 0 \\
 d35-x01-y01 & \ditto $B^{**}$ & 10 & 0 \\
 d36-x01-y01 & \ditto $D_{s1}^+$ & 10 & 0 \\
 d37-x01-y01 & \ditto $D_{s2}^+$ & 10 & 0 \\
 d38-x01-y03 & \ditto $p$ & 10 & 0 \\
 d39-x01-y03 & \ditto $\Lambda$ & 10 & 0 \\\hline

\end{tabular}
}
\caption{#1}
\label{table:longlisttwo}
\end{table}
}

\newcommand{\TABparamspytFix}{
\begin{table}[htbp]
\centering
\begin{tabular}{| c | c | c | c |}
    \hline
    Parameter & Range & Fixed & Default \\ \hline\hline
AlphaQCD        &     $[0.1000, 0.1417]$    &  x            & 0.1181    \\ 
IRcutoff (GeV) &     $[0.2002, 1.8014]$    &  x            & 1.0080    \\ 
SigmaPT (GeV)  &     $[0.000, 1.000]$      &  x            & 0.335     \\ 
aLund           &     $[0.20, 2.00]$        &  x            & 0.68      \\ 
bLund ($\text{GeV}^{-2}$) &     $[0.00, 2.00]$        &  x            & 0.98      \\ 
aExtraDiQuark   &     $[0.00, 2.00]$        &  \checkmark   & 0.97      \\ 
aExtraSQuark    &     $[0.00, 2.00]$        &  \checkmark   & 0.00      \\\hline
\end{tabular}
\caption{Parameters for the \texttt{Herwig7-P8} tune, their ranges and default values. The parameters marked as fixed are set to their default values for the tuning process. The quantities without units are dimensionless.}
\label{table:parameterP8}
\end{table}
}

\newcommand{\TABresultTuneHer}{
\begin{table}[htbp]
\centering
\begin{tabular}{| c | c | c | c |}
    \hline
    Parameter & Global & Marginal & Smallest \\ 
     & mode     &     mode              & 68\% interval \\ \hline\hline
AlphaQCD  & 0.115 & 0.115 &              [0.112, 0.118] \\
IRCutoff (GeV) & 0.879 & 0.755 &              [0.580, 1.020] \\
$m_{g}$(GeV) & 0.709 & 0.738 &              [0.700, 0.955] \\
$m_{s}$(GeV) & 0.353 & 0.375 &              [0.346, 0.470] \\
ClMax (GeV)& 2.591 & 4.025 &              [3.200, 4.750] \\
ClPow     & 0.823 & 0.910 &  [0.740, 1.260], [1.540, 2.260] \\
ClSmr     & 0.675 & 0.725 &              [0.480, 0.885] \\
PSplit    & 0.868 & 0.728 &              [0.615, 0.865] \\\hline
\end{tabular}
\caption{Results of the tune of the \texttt{Herwig7-H7} model. The values of the global and marginalized mode of the posterior samples as well as the smallest intervals containing 68\% of the probability are listed.}
\label{table:resultTuneHer}
\end{table}
}

\newcommand{\TABresultTunePytFix}{
\begin{table}[htbp]
\centering
\begin{tabular}{| c | c | c | c | c |}
    \hline
    Parameter & Global & Marginal & Smallest   & Fixed \\ 
    &          mode   & mode             &  68\% interval &  \\ \hline\hline 
AlphaQCD       & 0.120 & 0.120  & [0.117, 0.122] & x \\
IRCutOff (GeV) & 1.079 & 1.079  & [0.730, 1.390] & x \\
SigmaPT (GeV)  & 0.303 & 0.311  & [0.284, 0.336] & x \\
aLund          & 1.287 & 1.435  & [0.950, 1.760] & x \\
bLund ($\text{GeV}^{-2}$)         & 1.302 & 1.325  & [0.940, 1.720] & x \\
aExtraDiquark  & 0.970 & 0.970  & - & \checkmark  \\
aExtraSQuark   &   0.0 &   0.0  & - & \checkmark  \\\hline
\end{tabular}
\caption{Results of the tune of the \texttt{Herwig7-P8} model. The values of the global and marginalized mode of the posterior samples as well as the smallest intervals containing 68\% of the probability are listed. 
Fixed parameters are set to their mode value. The quantities without units are dimensionless.}
\label{table:resultTunePyt}
\end{table}
}

\newcommand{\TABcorrMode}{
\begin{table}[htbp]
\centering
\begin{tabular}{| c | c | c || c | c |}
    \hline
                & \multicolumn{4}{c|}{Correlation} \\  
                & \multicolumn{2}{c||}{$r = 0.0$} & \multicolumn{2}{c|}{$r = 0.9$} \\ \hline 
    Parameter & Mode & $\sigma$ & Mode & $\sigma$ \\ \hline\hline
 AlphaQCD & 0.114 & 0.0032  & 0.115  &  0.0018 \\
     IRCutoff (GeV) & 0.810 &   0.221 & 0.730  &    0.126 \\
     $m_{g}$ (GeV) & 0.716 &   0.129 & 0.780  &     0.106 \\
   $m_{s}$ (GeV) & 0.357 &  0.062  & 0.390  &   0.050 \\
 ClMax (GeV) & 2.543 &   0.872 &  2.081 &     0.742 \\
    ClPow & 0.800 &   0.560 & 0.667  &    0.545 \\
    ClSmr & 0.662 &   0.194 & 0.461  &    0.148 \\
    PSplit & 0.916 &   0.128 &  1.050 &    0.101 \\\hline
\end{tabular}
\caption{The mode and standard deviation of the tuned parameters for the tune of \texttt{Herwig7-H7}. Results are shown without correlation and for a correlation of $r = 0.9$. The quantities without units are dimensionless.}
\label{table:corrMode}
\end{table}
}

\newcommand{\TABweightHerBig}{
\setlength{\tabcolsep}{0.35em}
\begin{table}[htbp]\centering
\begin{tabular}{| c | c | c || c | c || c | c |}\hline
          & \multicolumn{6}{c|}{Weighting scheme} \\
          & \multicolumn{2}{c||}{none} & \multicolumn{2}{c||}{$w_1$} & \multicolumn{2}{c|}{$w_2$} \\\hline
Parameter & Mode & $\sigma$ & Mode & $\sigma$ & Mode & $\sigma$ \\ \hline\hline
 AlphaQCD & 0.115 &   0.003  & 0.113 & 0.004 & 0.115 & 0.003 \\
 IRCutoff (GeV)  & 0.879 &   0.223  & 0.859 & 0.245 & 0.837 & 0.214 \\
     $m_{g}$ (GeV) & 0.709 &   0.128  & 0.706 & 0.136 & 0.708 & 0.13  \\
     $m_{s}$ (GeV) & 0.353 &   0.062  & 0.352 & 0.066 & 0.346 & 0.063 \\
    ClMax (GeV) & 2.591 &   0.871  & 3.187 & 0.911 & 3.761 & 0.832 \\
    ClPow & 0.823 &   0.561  & 0.847 & 0.567 & 2.147 & 0.541 \\
    ClSmr & 0.675 &   0.193  & 0.501 & 0.228 & 0.806 & 0.213 \\
   PSplit & 0.868 &   0.130  & 0.867 & 0.141 & 0.776 & 0.137 \\\hline
\end{tabular}
\caption{The mode and standard deviation of the tuned parameters for the tune of \texttt{Herwig7-H7} using different weighting schemes. The quantities without units are dimensionless.}
\label{table:weightHerBig}
\end{table}
}

\newcommand{\TABweightPytBig}{
\setlength{\tabcolsep}{0.35em}
\begin{table}[htbp]\centering
\begin{tabular}{| c | c | c || c | c || c | c |}\hline

          & \multicolumn{6}{c|}{Weighting scheme} \\
          & \multicolumn{2}{c||}{none} & \multicolumn{2}{c||}{$w_1$} & \multicolumn{2}{c|}{$w_2$} \\\hline
Parameter & Mode & $\sigma$ & Mode & $\sigma$ & Mode & $\sigma$ \\ \hline\hline
AlphaQCD         & 0.120    & 0.003  & 0.120   & 0.004 & 0.120  & 0.003 \\
IRCutOff (GeV) & 1.079    & 0.313  & 1.135   & 0.363 & 1.115  & 0.339 \\
aLund            & 1.287    & 0.376  & 1.380   & 0.416 & 1.381  & 0.396 \\
bLund ($\text{GeV}^{-2}$) & 1.302    & 0.359  & 1.369   & 0.379 & 1.393  & 0.363 \\
aExtraDiquark    &  0.97    &  -     &  0.97   & -     &  0.97  & -     \\
aExtraSQuark     &   0.0    &  -     &   0.0   & -     &   0.0  & -     \\
SigmaPT (GeV)    & 0.303    & 0.026  & 0.304   & 0.031 & 0.302  & 0.027 \\\hline
\end{tabular}
        \caption{The mode and standard deviation of the tuned parameters for the tune of \texttt{Herwig7-P8} using different weighting schemes. The quantities without units are dimensionless.}
        \label{table:weightPytBig}
\end{table}
}

%% file: BayesHerwigTune-fig.tex
\makeatletter
\def\ifabsgreater #1#2{\ifpdfabsnum 
\dimexpr#1pt>\dimexpr#2pt\relax
\expandafter\@firstoftwo\else\expandafter\@secondoftwo\fi }
\makeatother

\newcommand{\FIGone}[2]{
\begin{figure}[htb]\centering
\includegraphics[width=0.49\linewidth,height=0.27\linewidth]{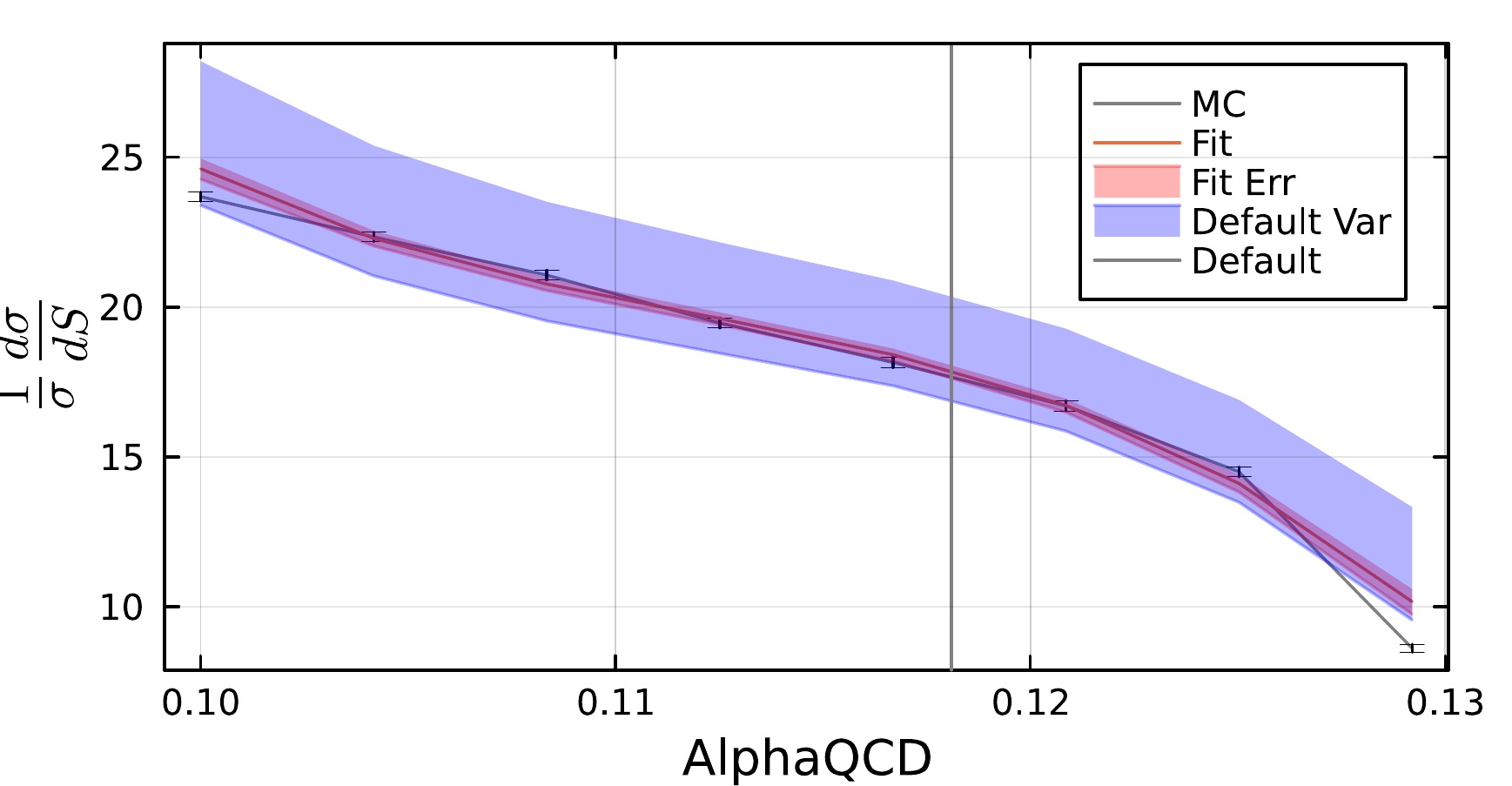}
\includegraphics[width=0.49\linewidth,height=0.27\linewidth]{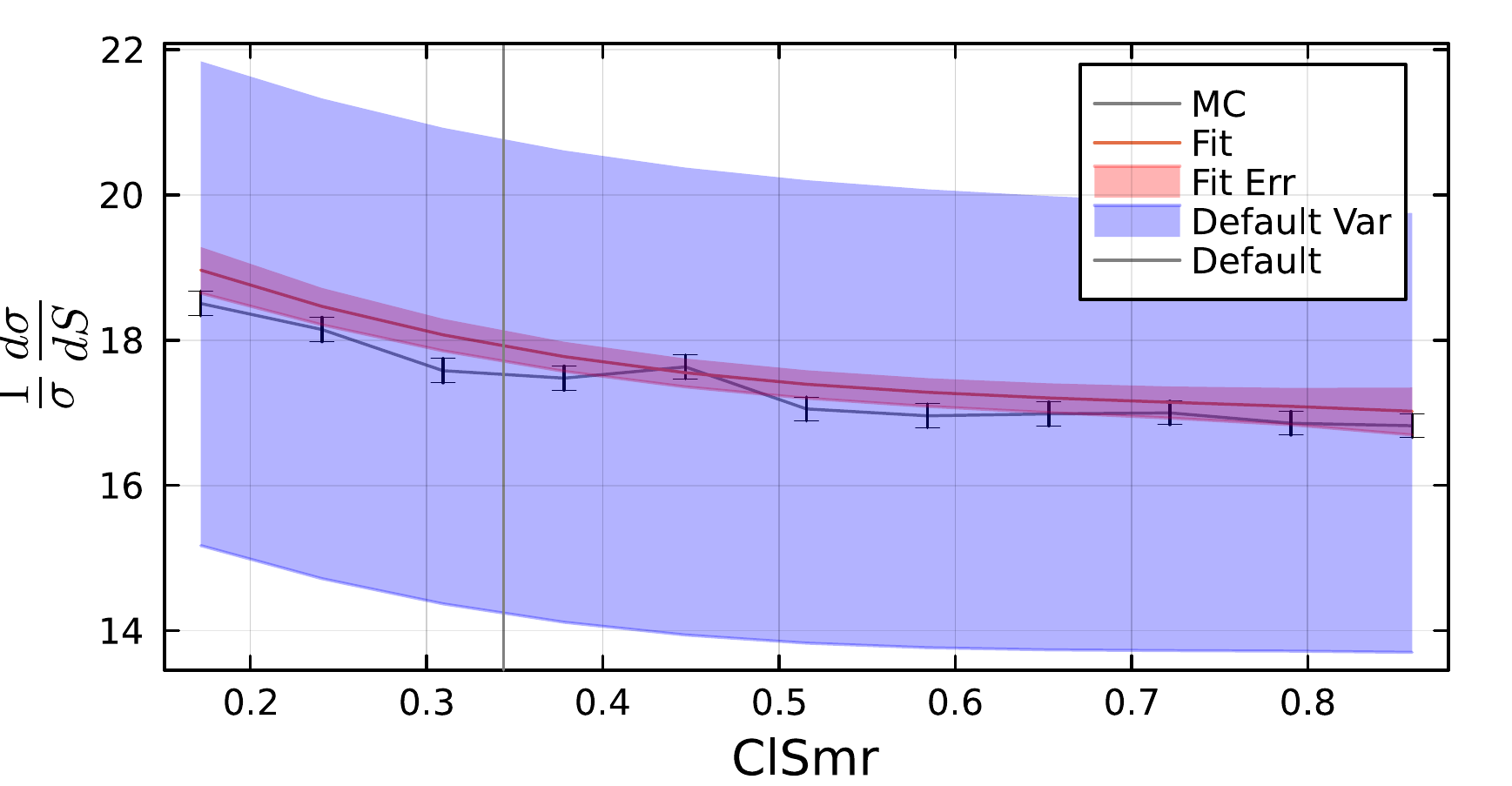}\\
\includegraphics[width=0.49\linewidth,height=0.27\linewidth]{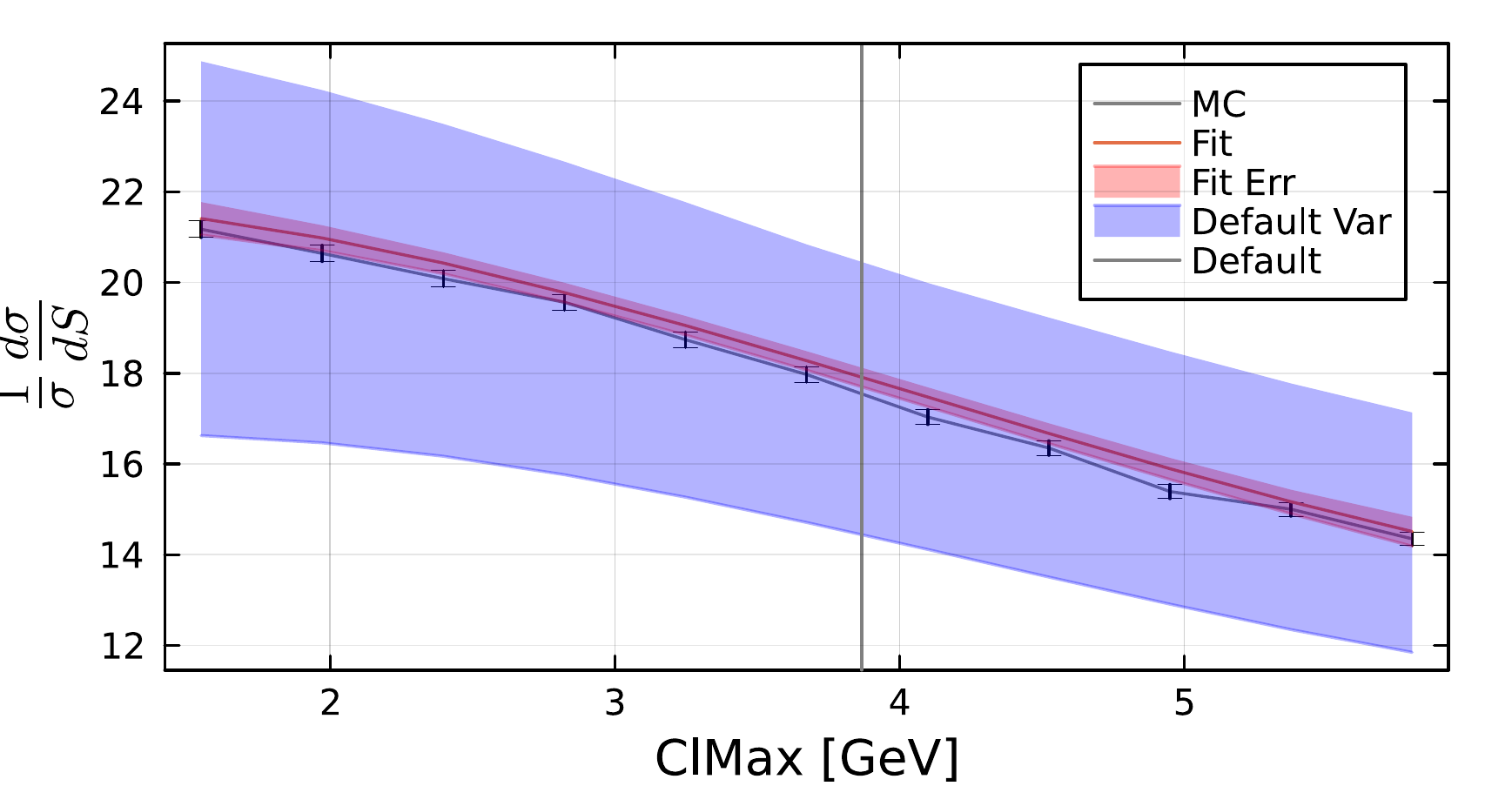}\includegraphics[width=0.49\linewidth,height=0.27\linewidth]{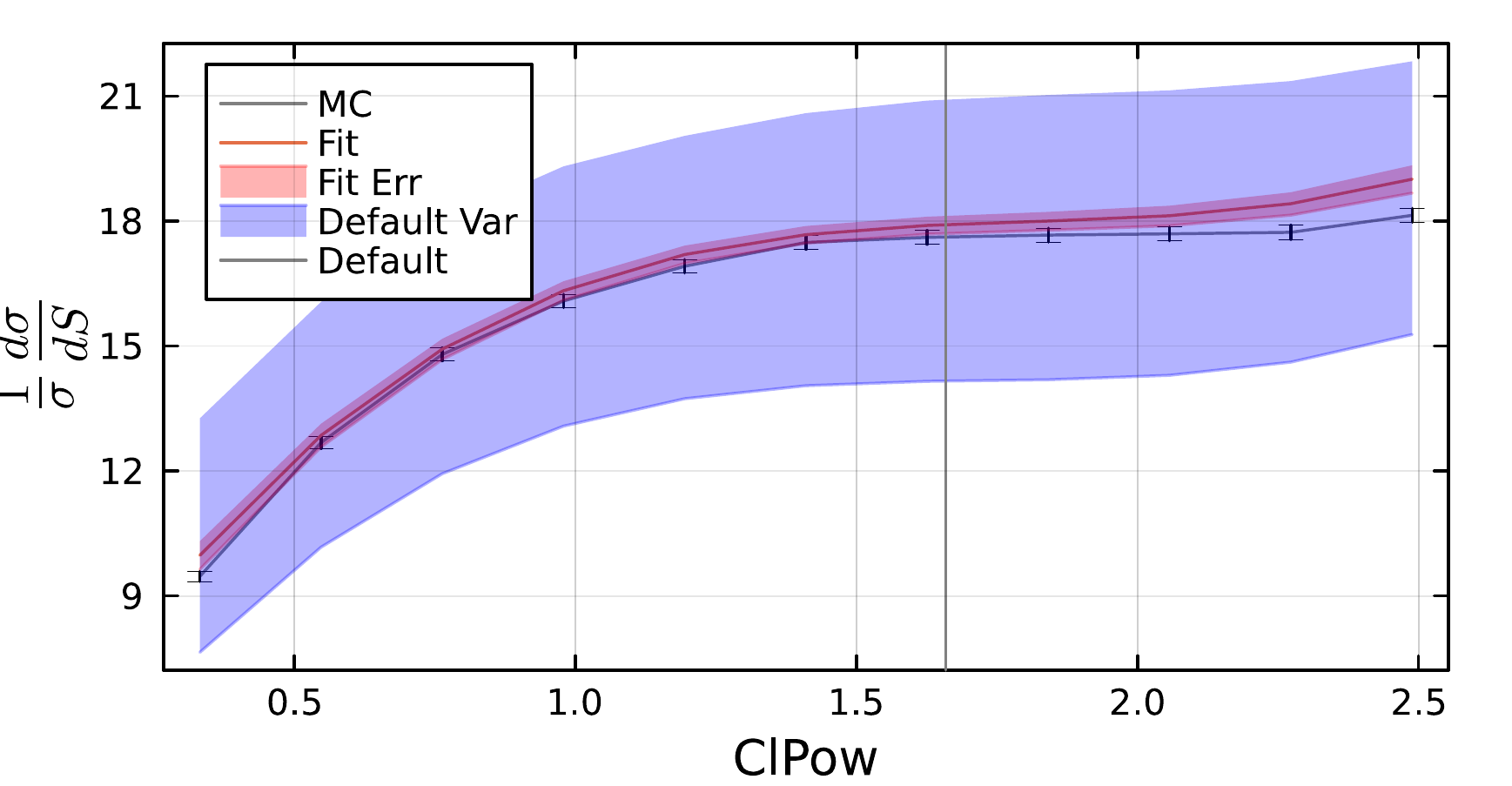}
\caption{#1}
\label{#2}
\end{figure}
}

\newcommand{\FIGtwo}[2]{
\begin{figure}[htbp]\centering
\includegraphics[width=0.7\linewidth,height=0.35\linewidth]{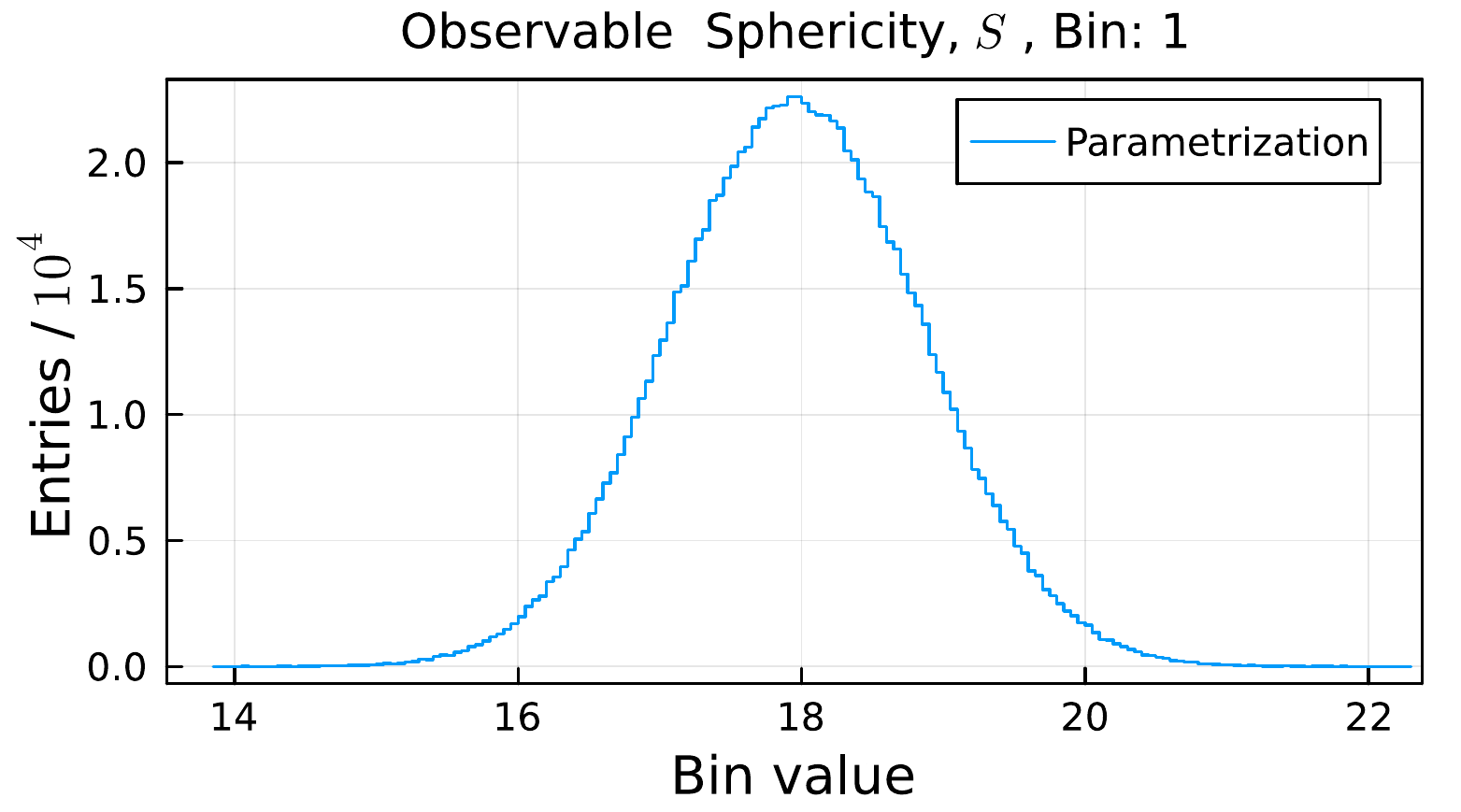}\\
\includegraphics[width=0.7\linewidth,height=0.35\linewidth]{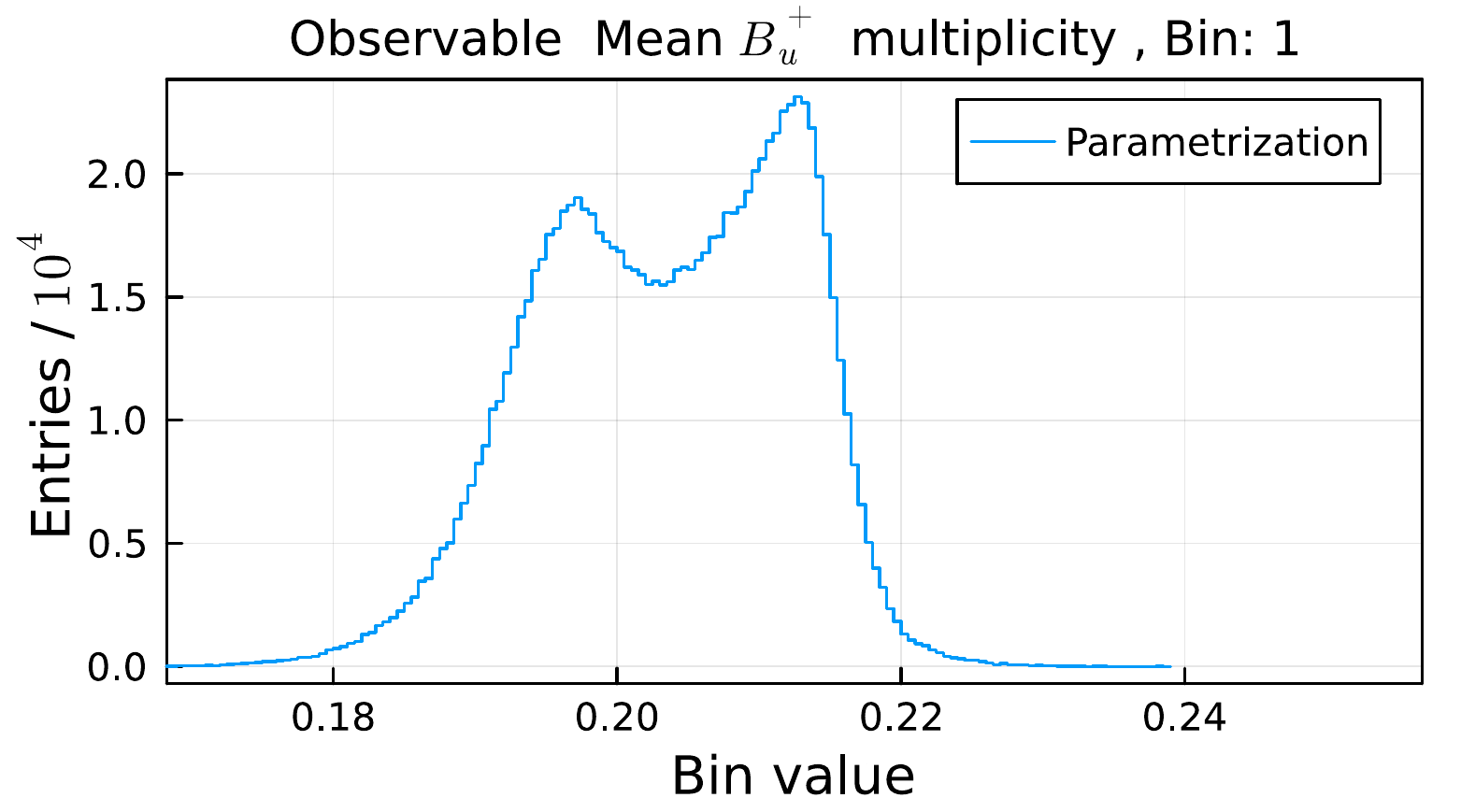}
\caption{#1}
\label{#2}
\end{figure}
}

\newcommand{\FIGthree}[2]{ 
\begin{figure}[htbp]\centering
\includegraphics[width=1.0\linewidth,height=1.0\linewidth]{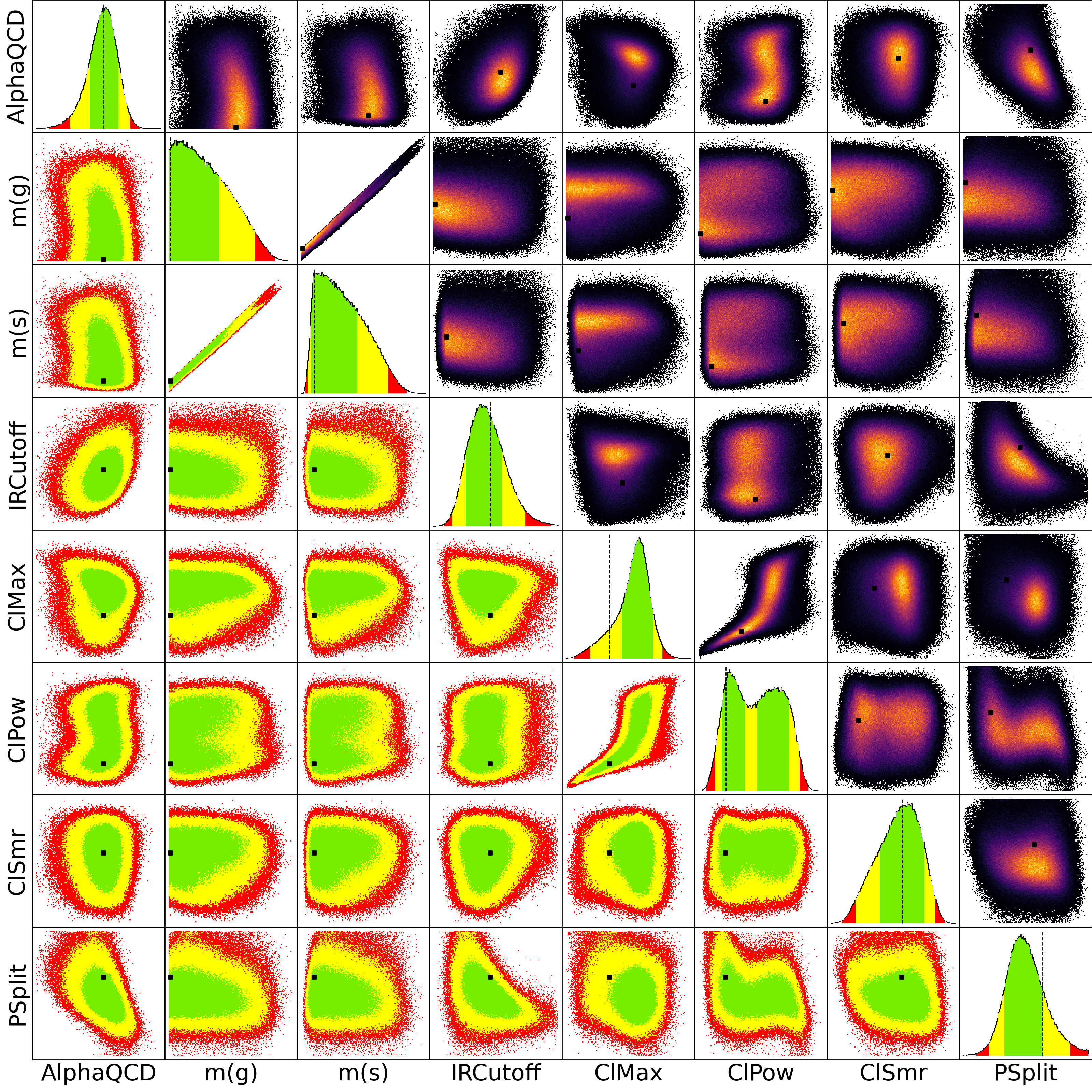}
\caption{#1} 
 \label{#2}
\end{figure}
}

\newcommand{\FIGfour}[2]{
\begin{figure}[htbp]\centering
\includegraphics[width=0.49\linewidth,height=0.24\linewidth]{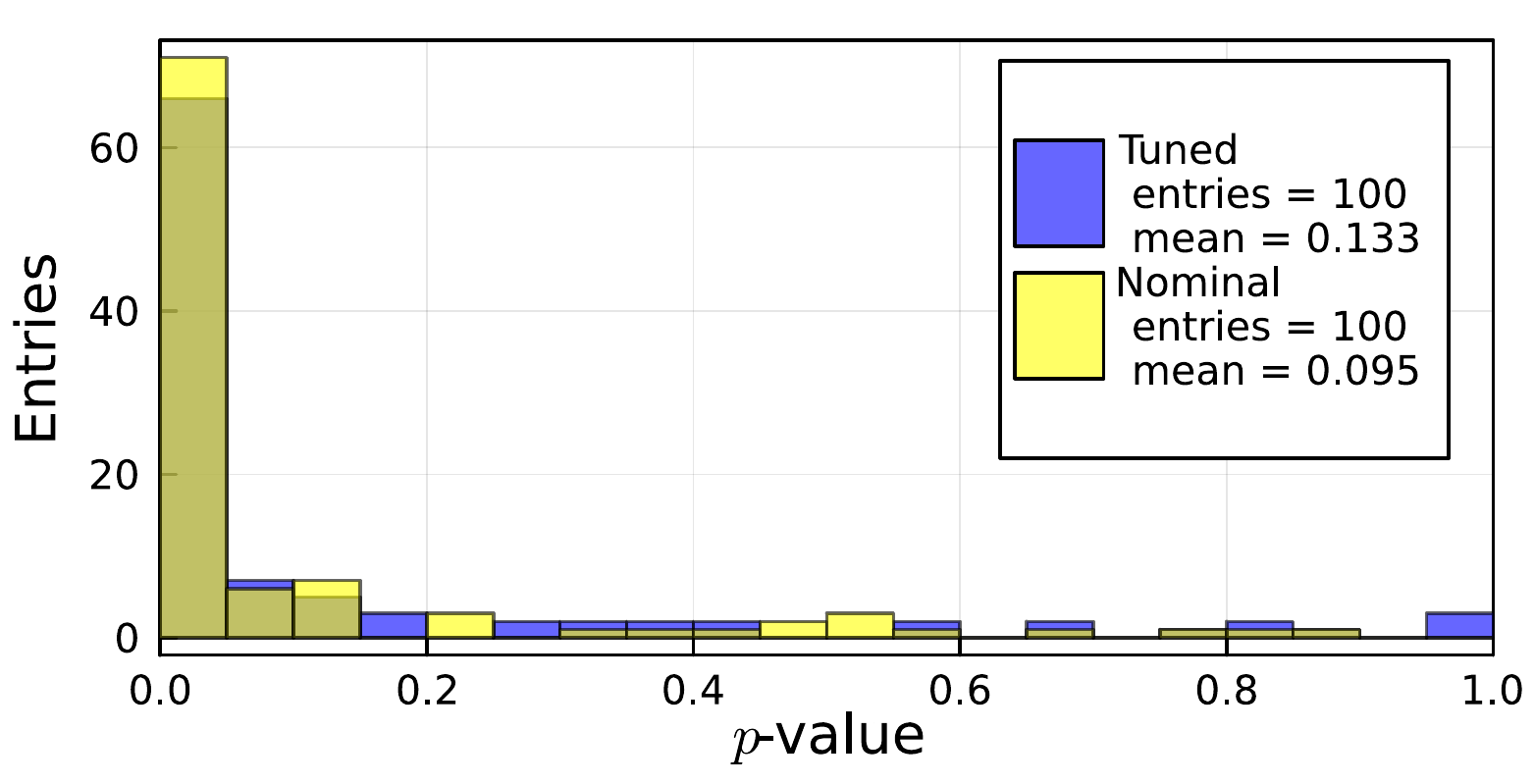}
\includegraphics[width=0.49\linewidth,height=0.24\linewidth]{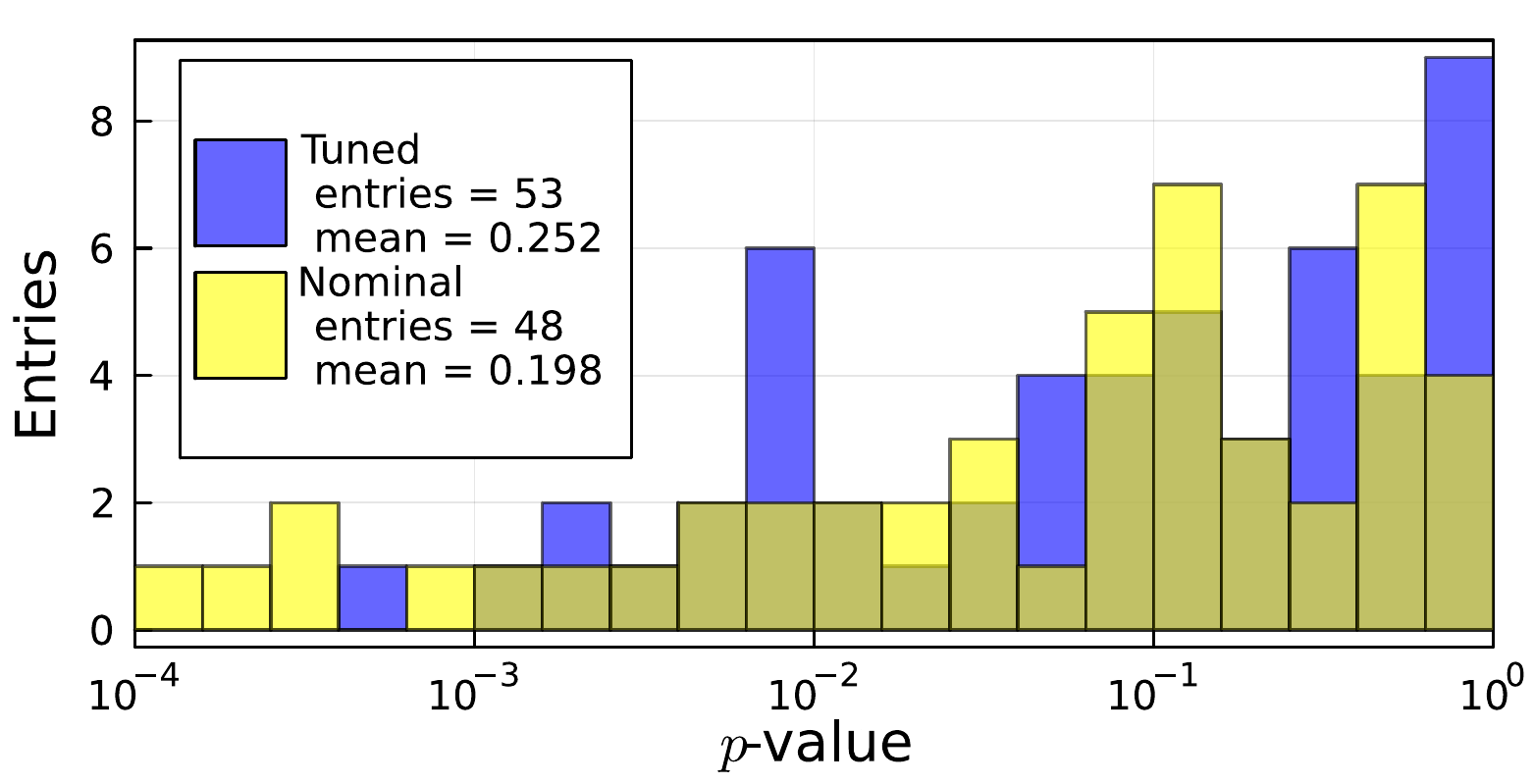}
\caption{#1}
\label{#2}
\end{figure}
}

\newcommand{\FIGfive}[2]{
\begin{figure}[!ht]\centering
\includegraphics[width=0.49\linewidth,height=0.49\linewidth]{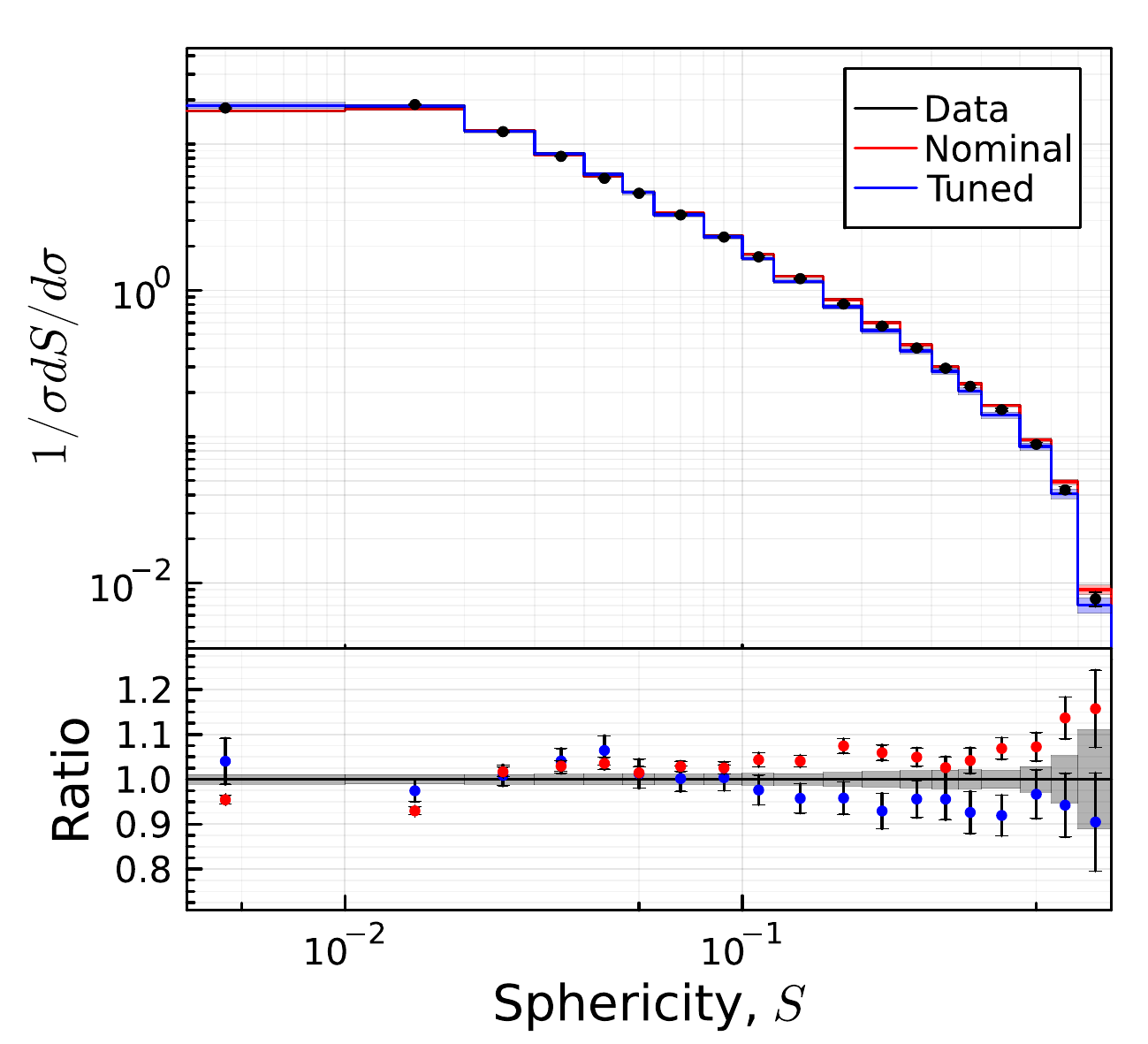}
\includegraphics[width=0.49\linewidth,height=0.49\linewidth]{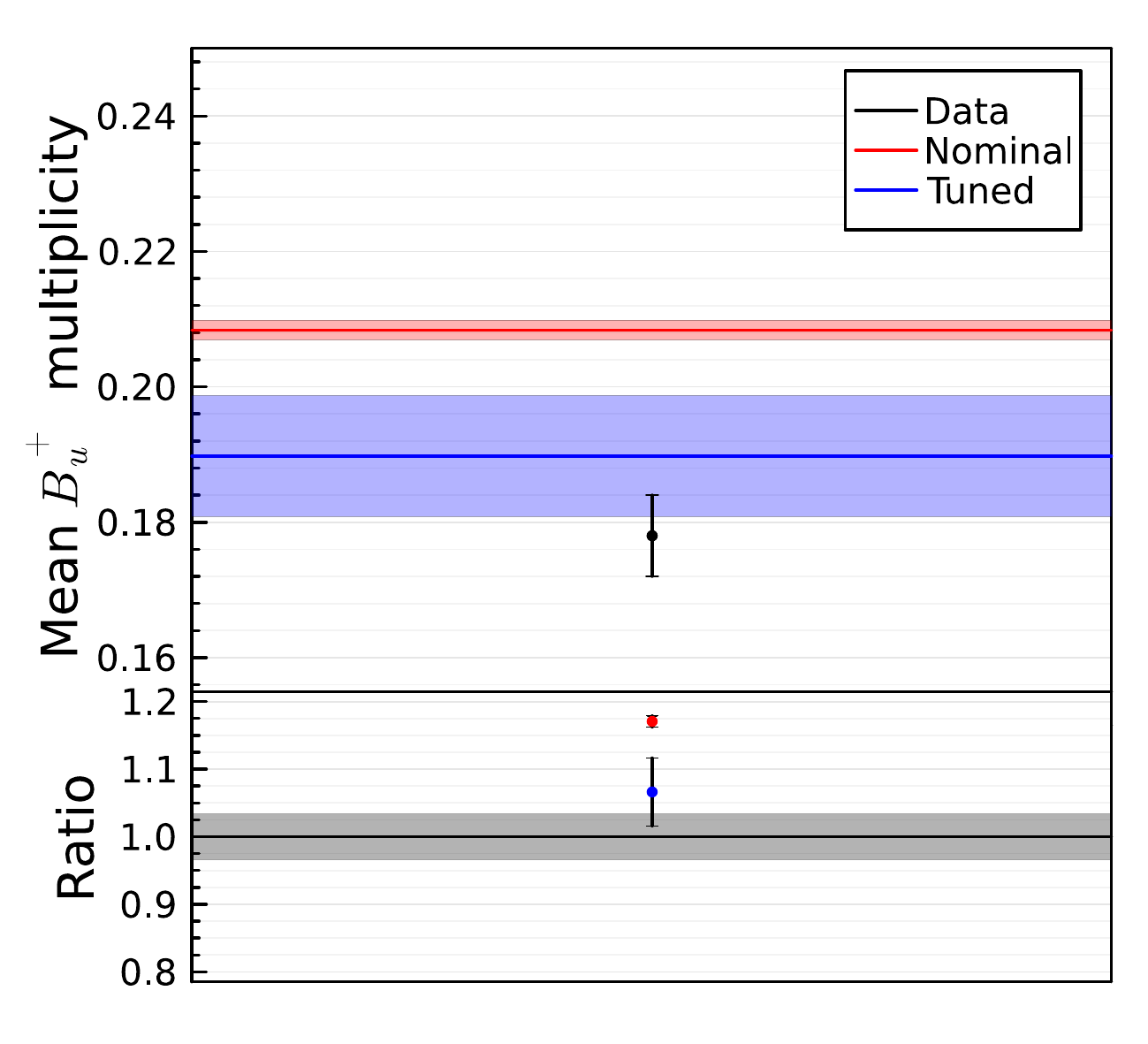}
\caption{#1}
\label{#2}
\end{figure}
}

\newcommand{\FIGsix}[2]{ 
\begin{figure}[htbp]\centering
\centering
\includegraphics[width=1.0\linewidth,height=1.0\linewidth]{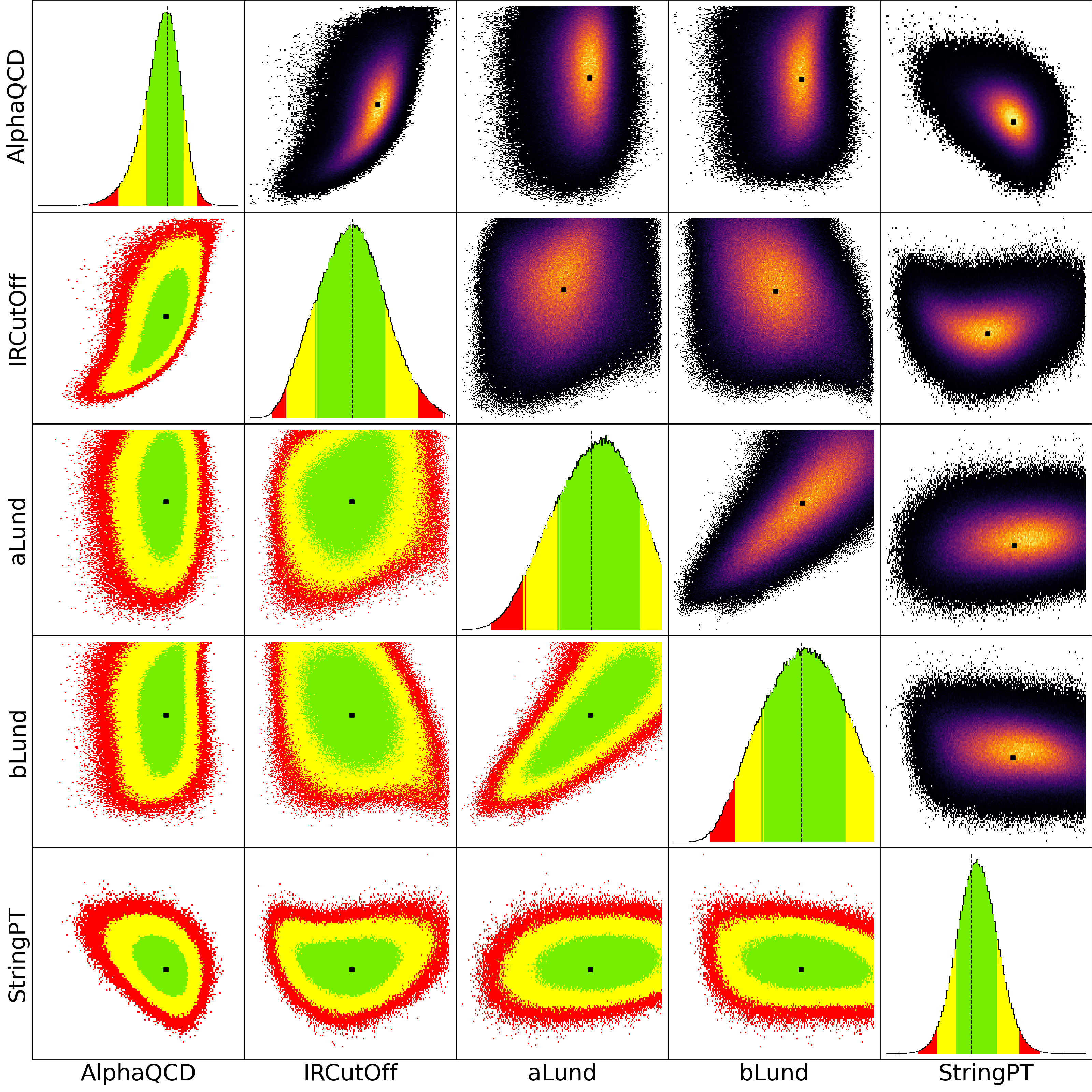}
\caption{#1} 
\label{#2}
\end{figure}
}

\newcommand{\FIGseven}[2]{
\begin{figure}[htbp]\centering
\includegraphics[width=0.49\linewidth,height=0.24\linewidth]{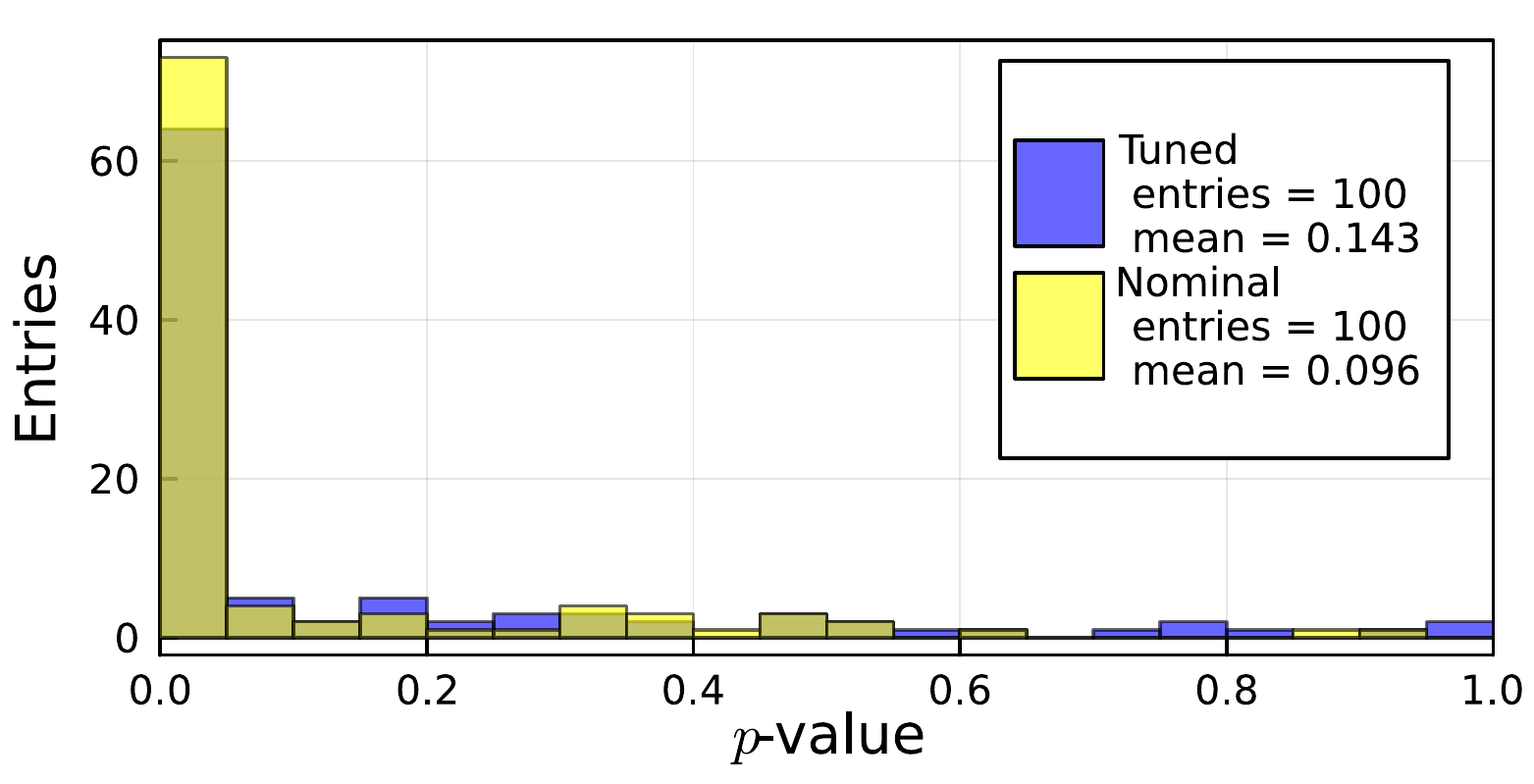}
\includegraphics[width=0.49\linewidth,height=0.24\linewidth]{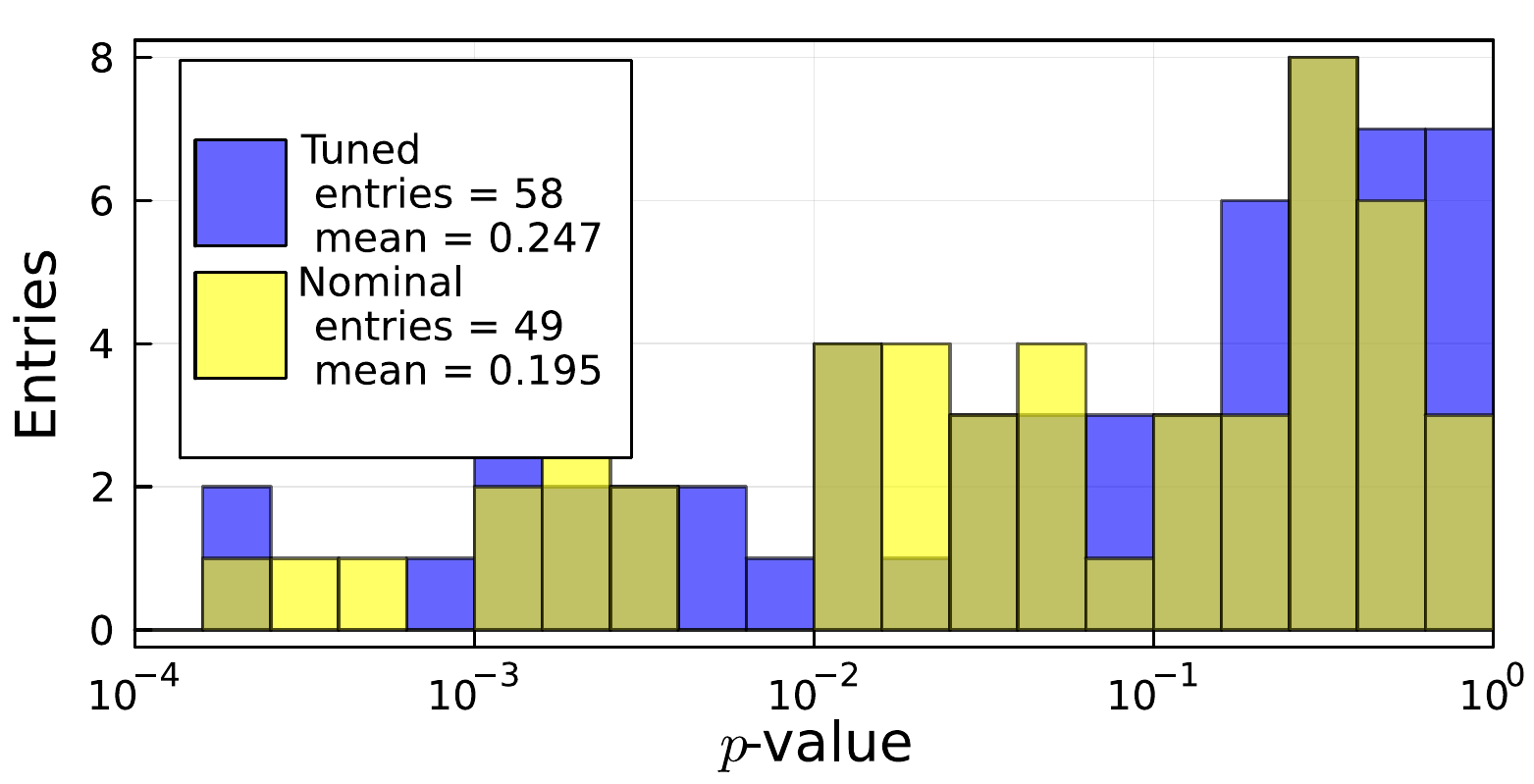}
\caption{#1}
\label{#2}
\end{figure}
}

\newcommand{\FIGeight}[2]{
\begin{figure}[htbp]\centering
\includegraphics[width=0.49\linewidth,height=0.490\linewidth]{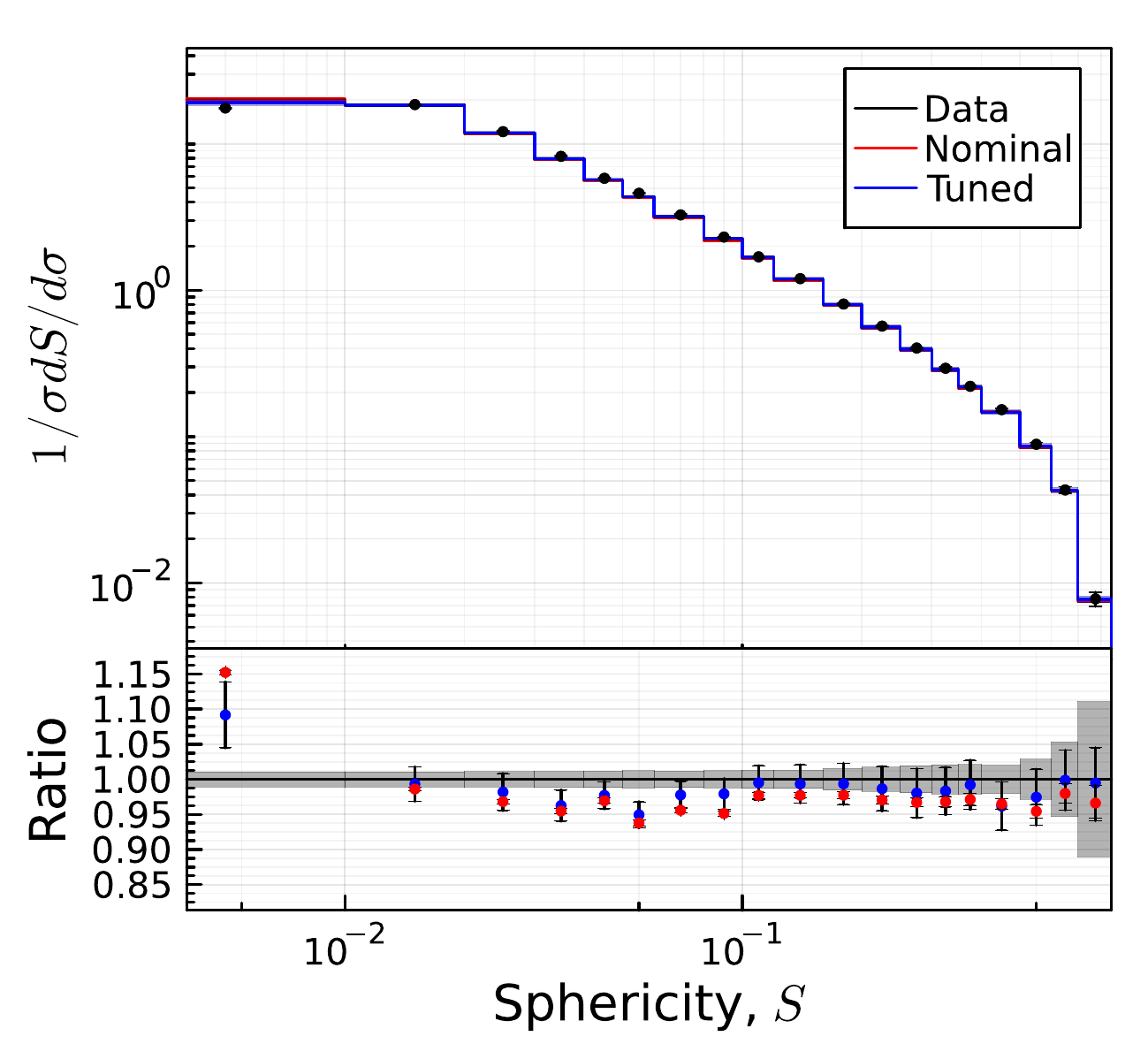}
\includegraphics[width=0.49\linewidth,height=0.490\linewidth]{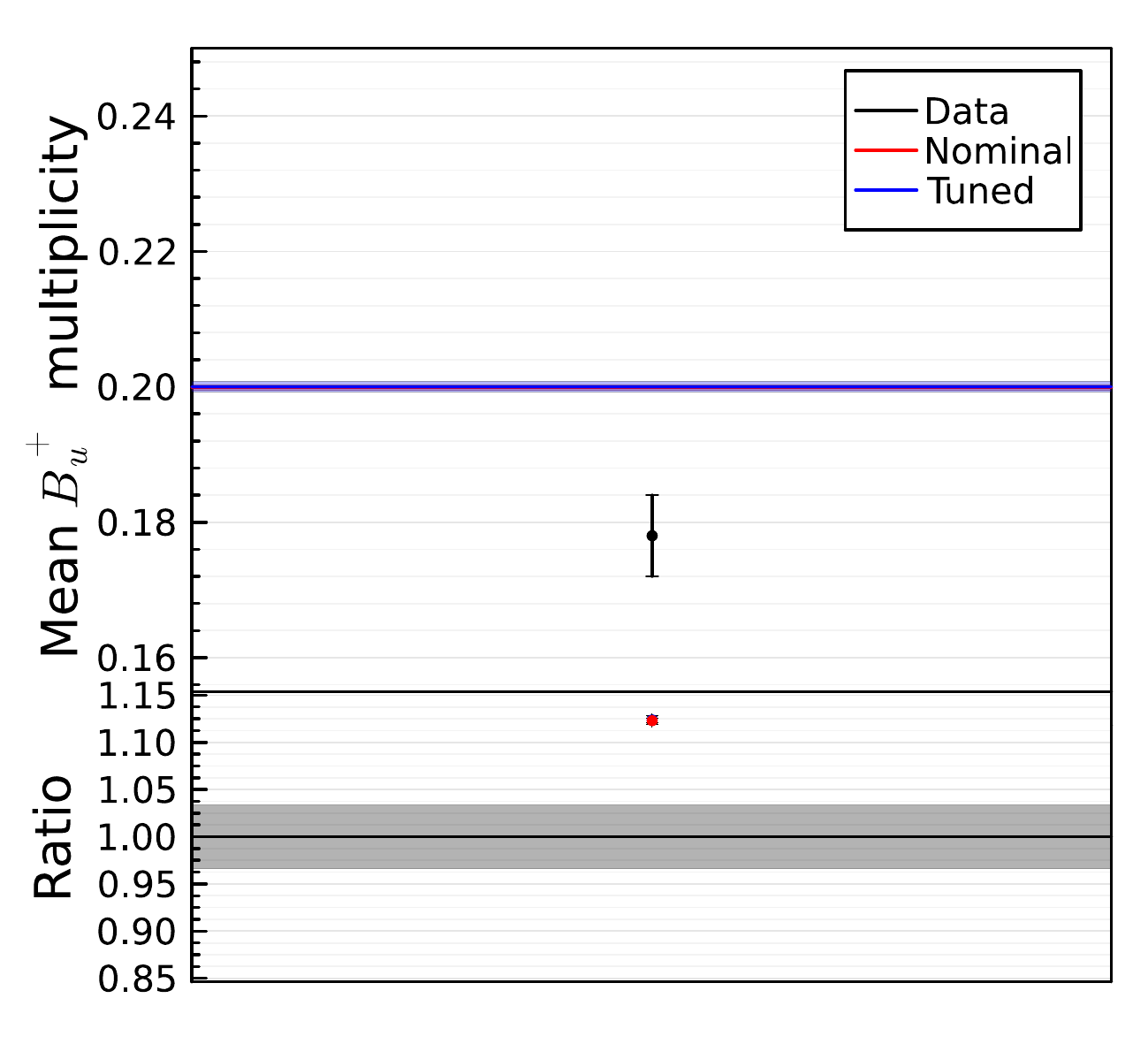}
\caption{#1}
\label{#2}
\end{figure}
}

\newcommand{\FIGnine}[2]{
\begin{figure}[htbp]\centering
\includegraphics[width=0.49\linewidth,height=0.28\linewidth]{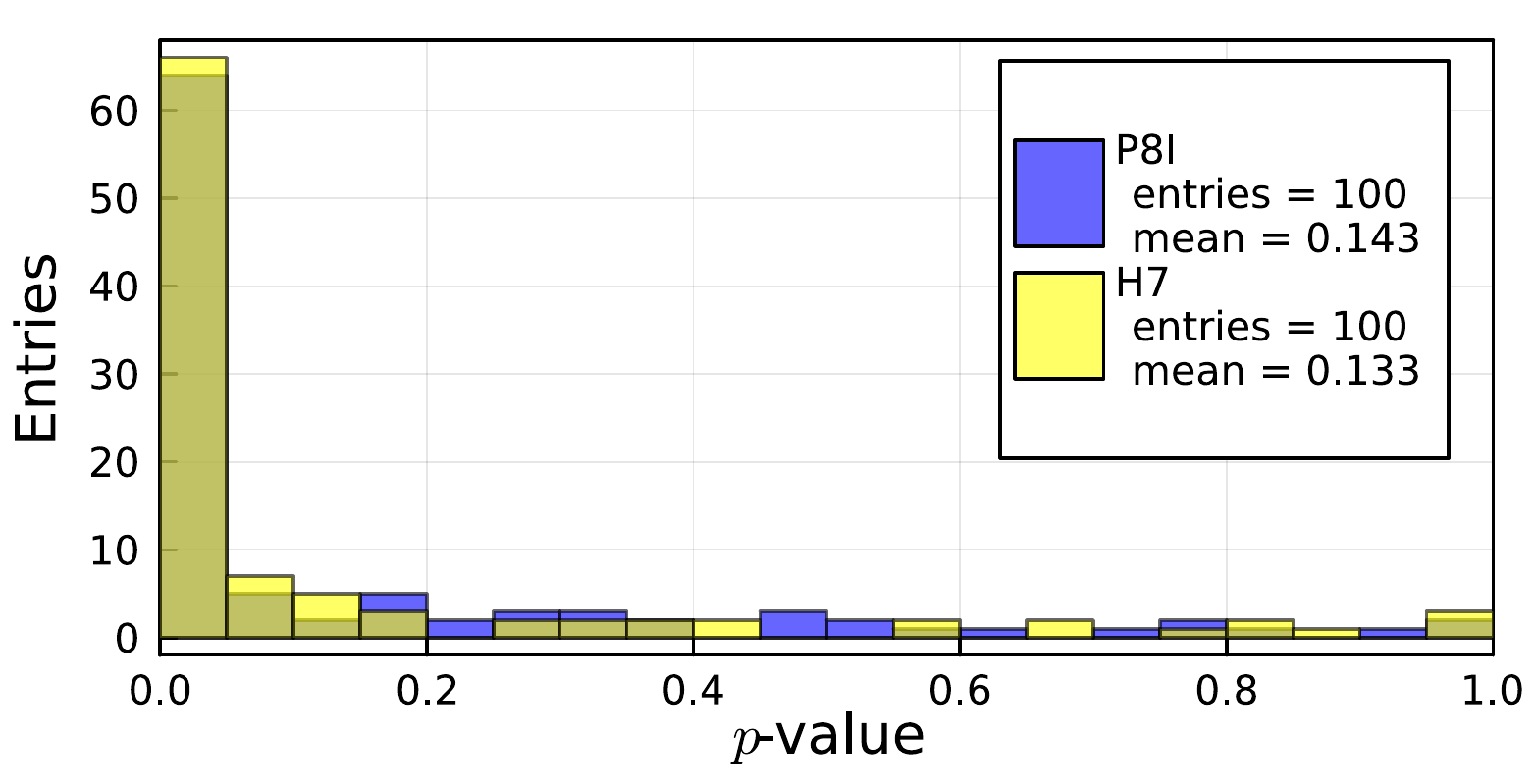}
\includegraphics[width=0.49\linewidth,height=0.28\linewidth]{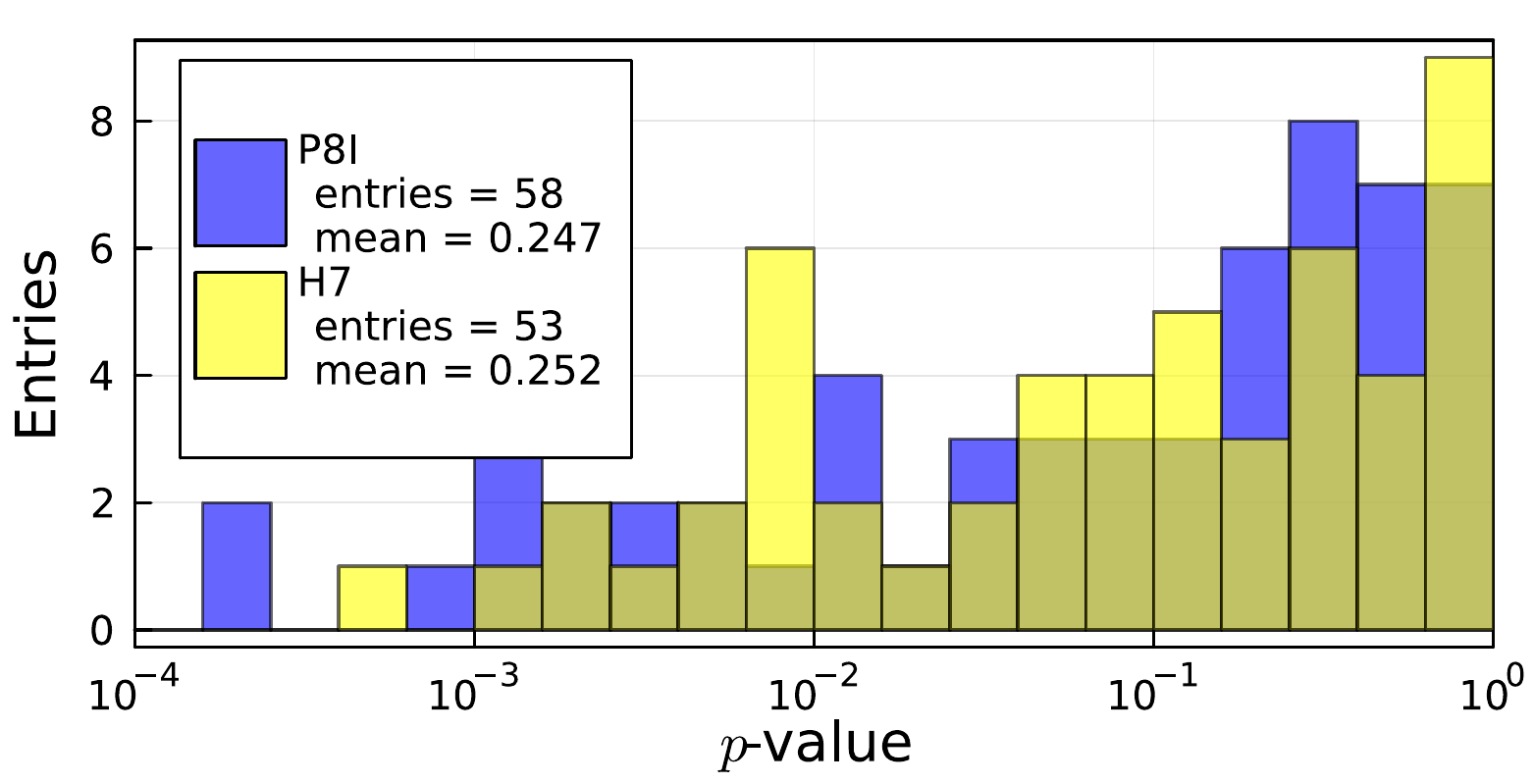}
\caption{#1}
\label{#2}
\end{figure}
}

\newcommand{\FIGten}[2]{
\begin{figure}[htbp]\centering
\includegraphics[width=0.49\linewidth,height=0.490\linewidth]{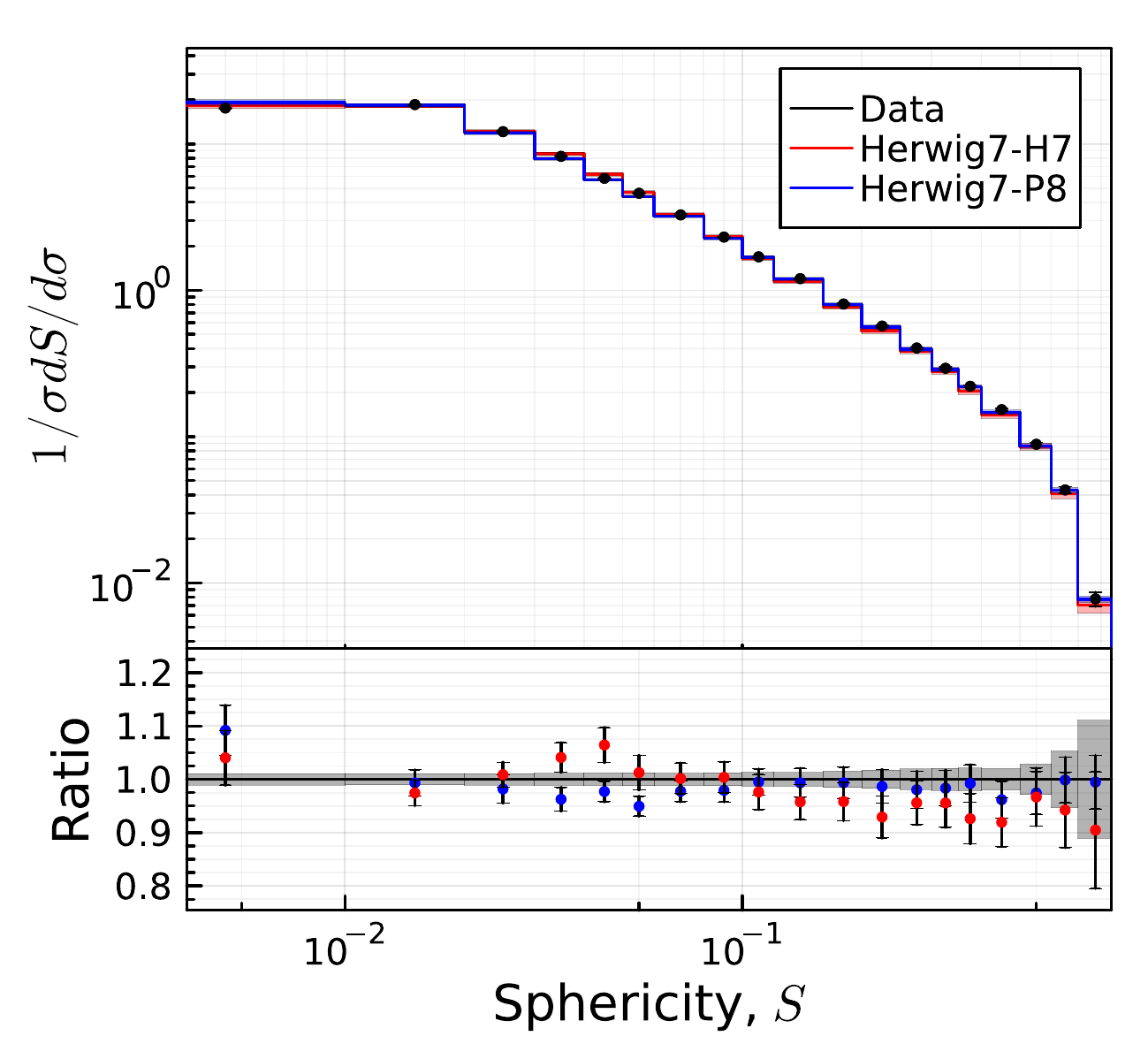}
\includegraphics[width=0.49\linewidth,height=0.490\linewidth]{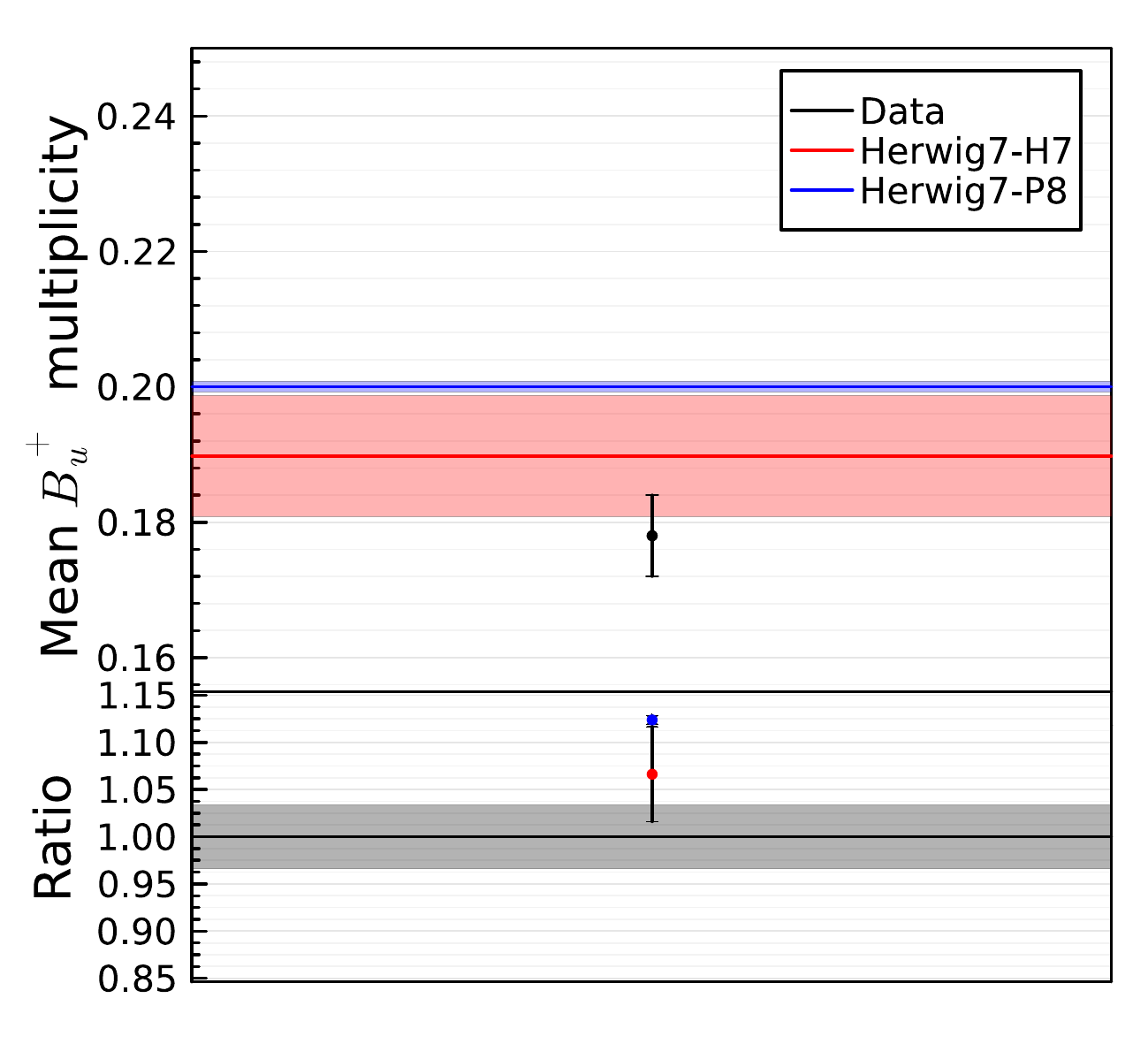}
\caption{#1}
\label{#2}
\end{figure}
}

\newcommand{\FIGeleven}[2]{
\begin{figure}[htbp]\centering
\centering
\includegraphics[width=0.7\linewidth]{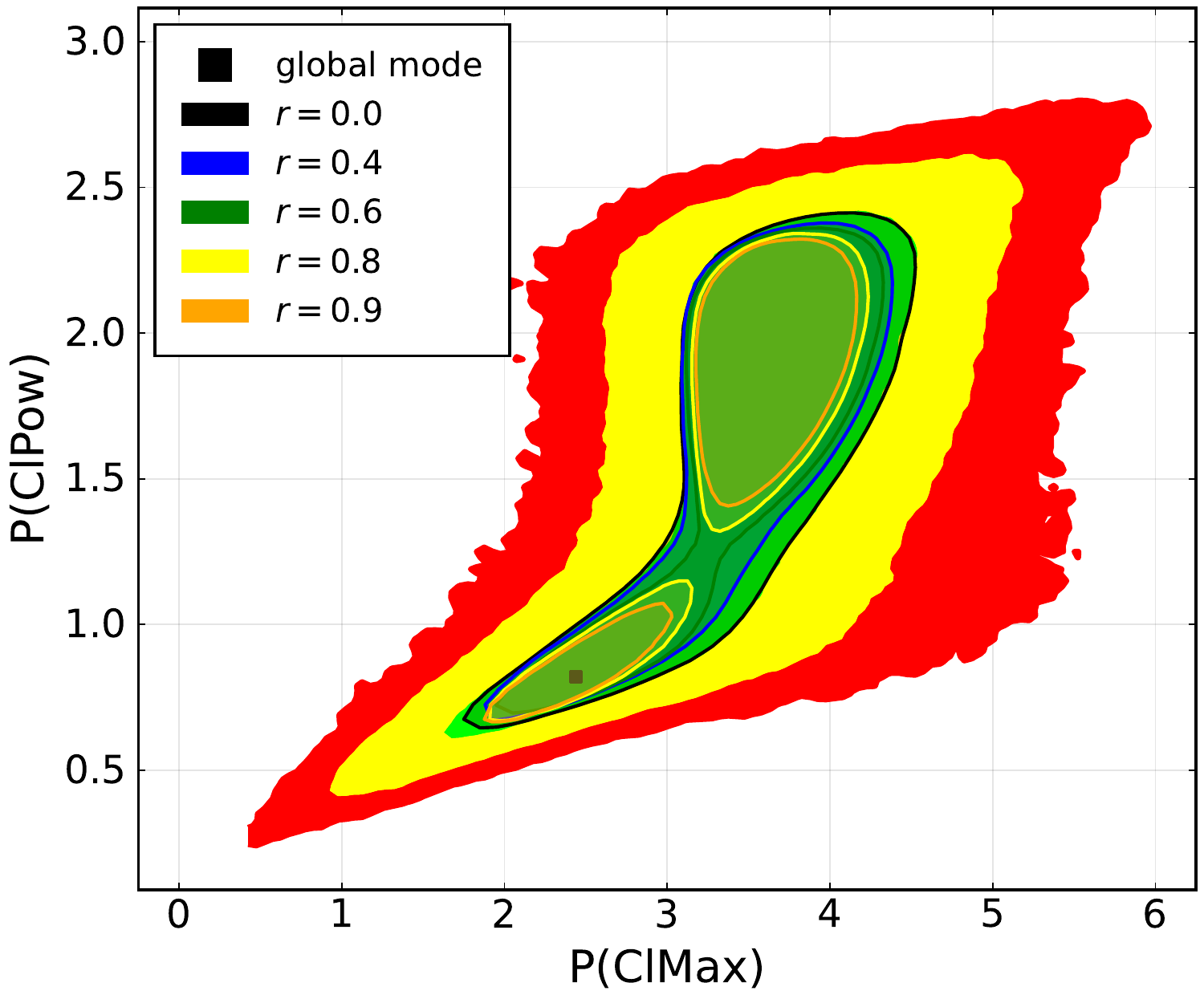}
\caption{#1}
\label{#2}
\end{figure}
}

\newcommand{\FIGtwelve}[2]{
\begin{figure}[htbp]\centering
\includegraphics[width=0.49\linewidth,height=0.28\linewidth]{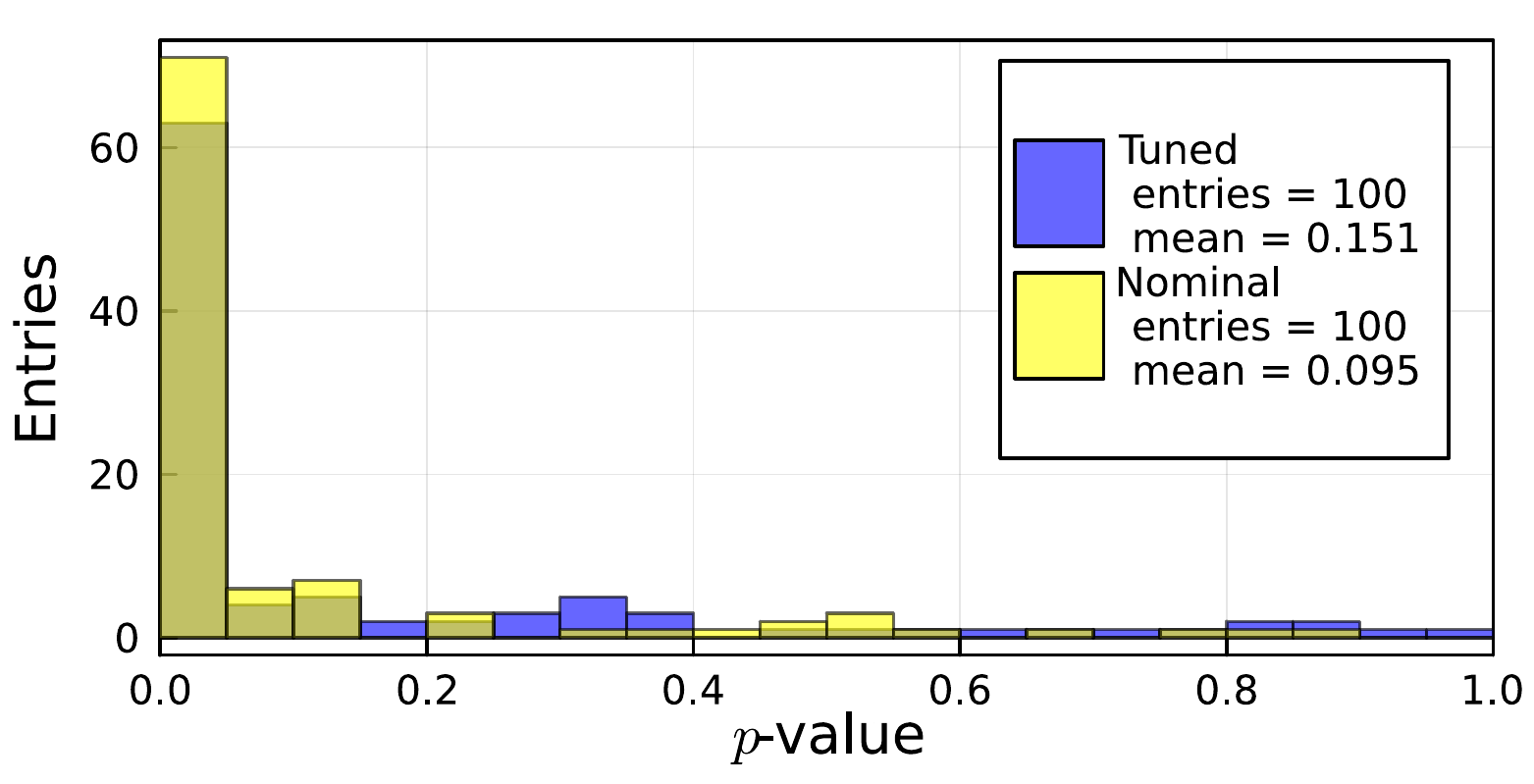}
\includegraphics[width=0.49\linewidth,height=0.28\linewidth]{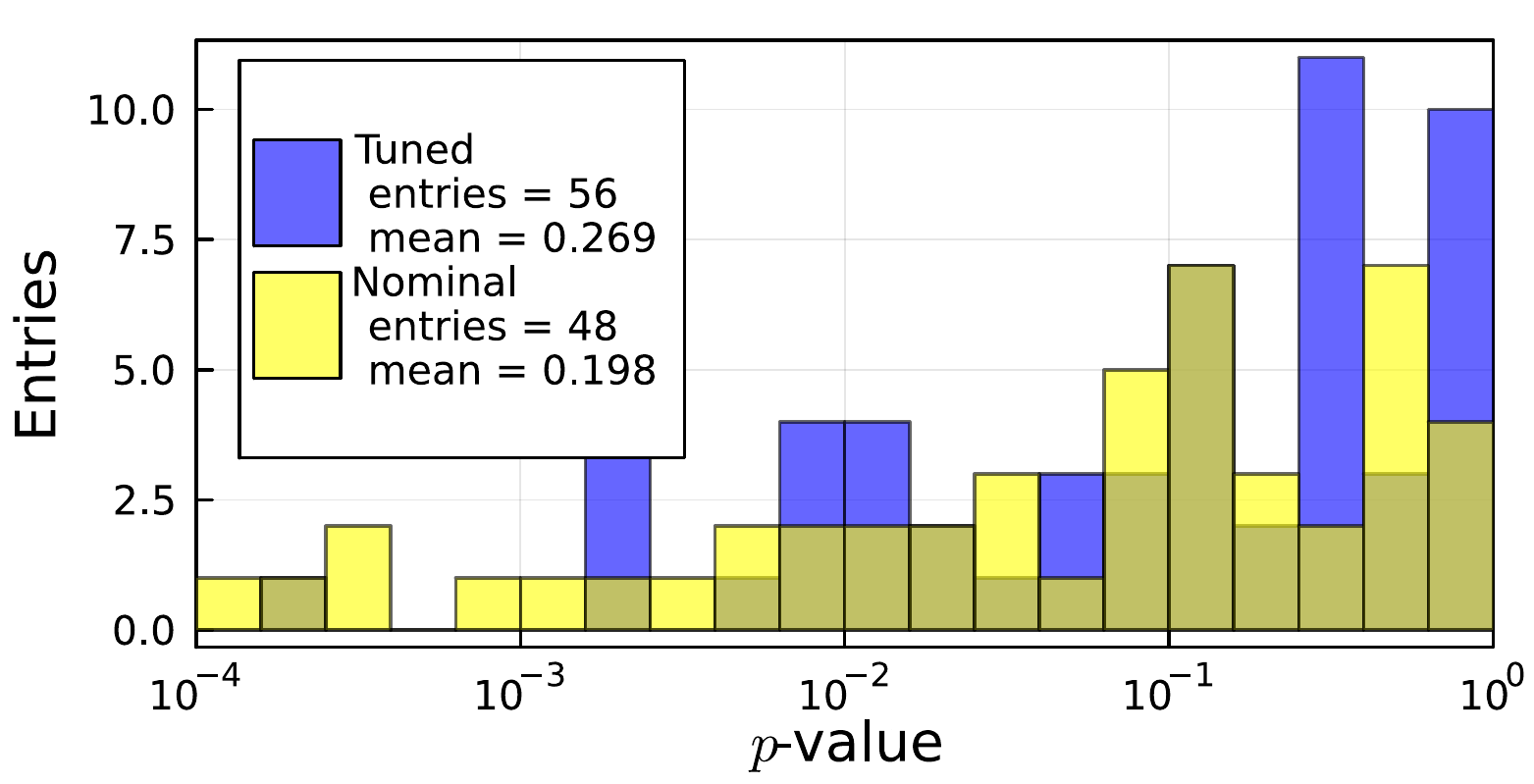}\\
\includegraphics[width=0.49\linewidth,height=0.28\linewidth]{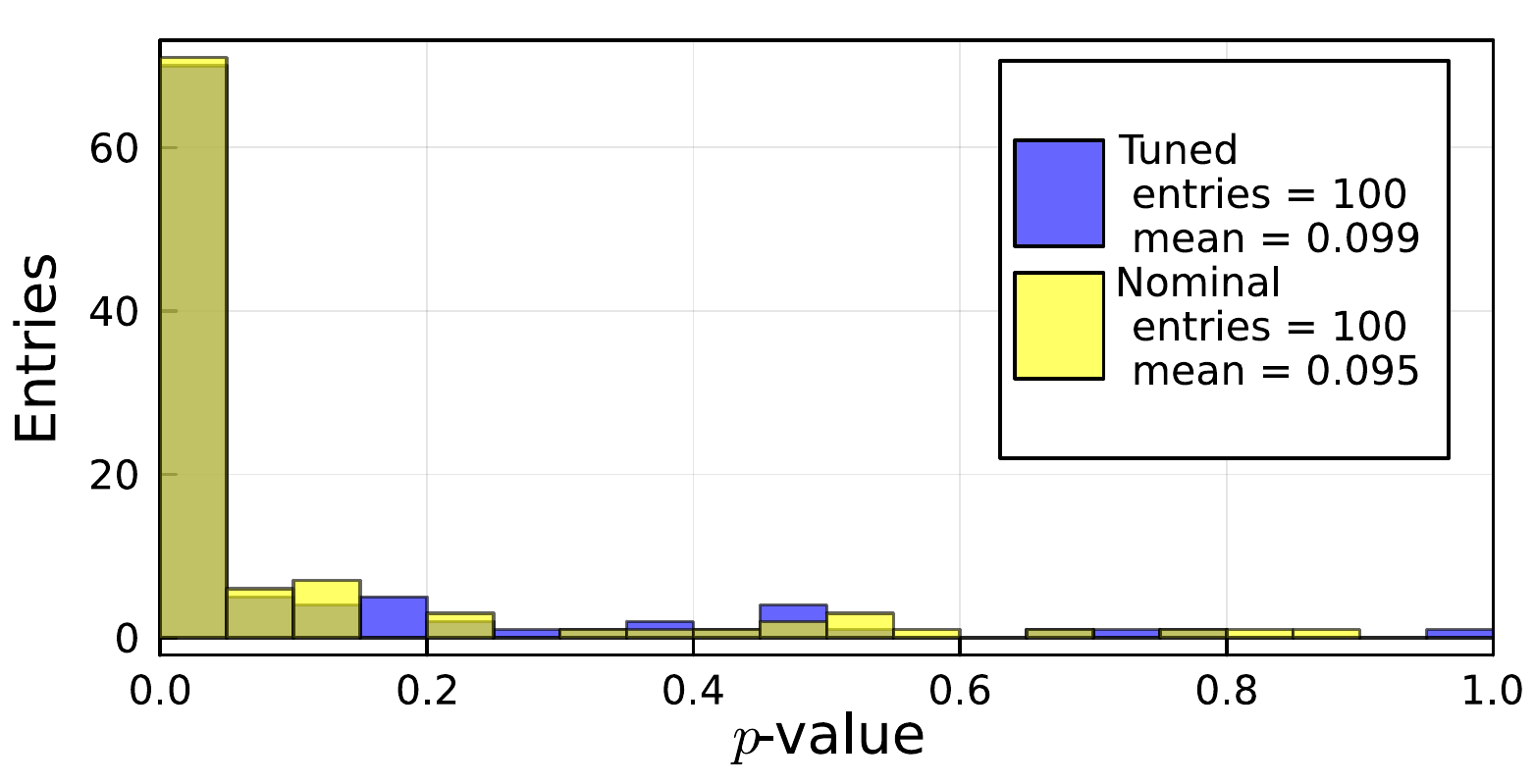}
\includegraphics[width=0.49\linewidth,height=0.28\linewidth]{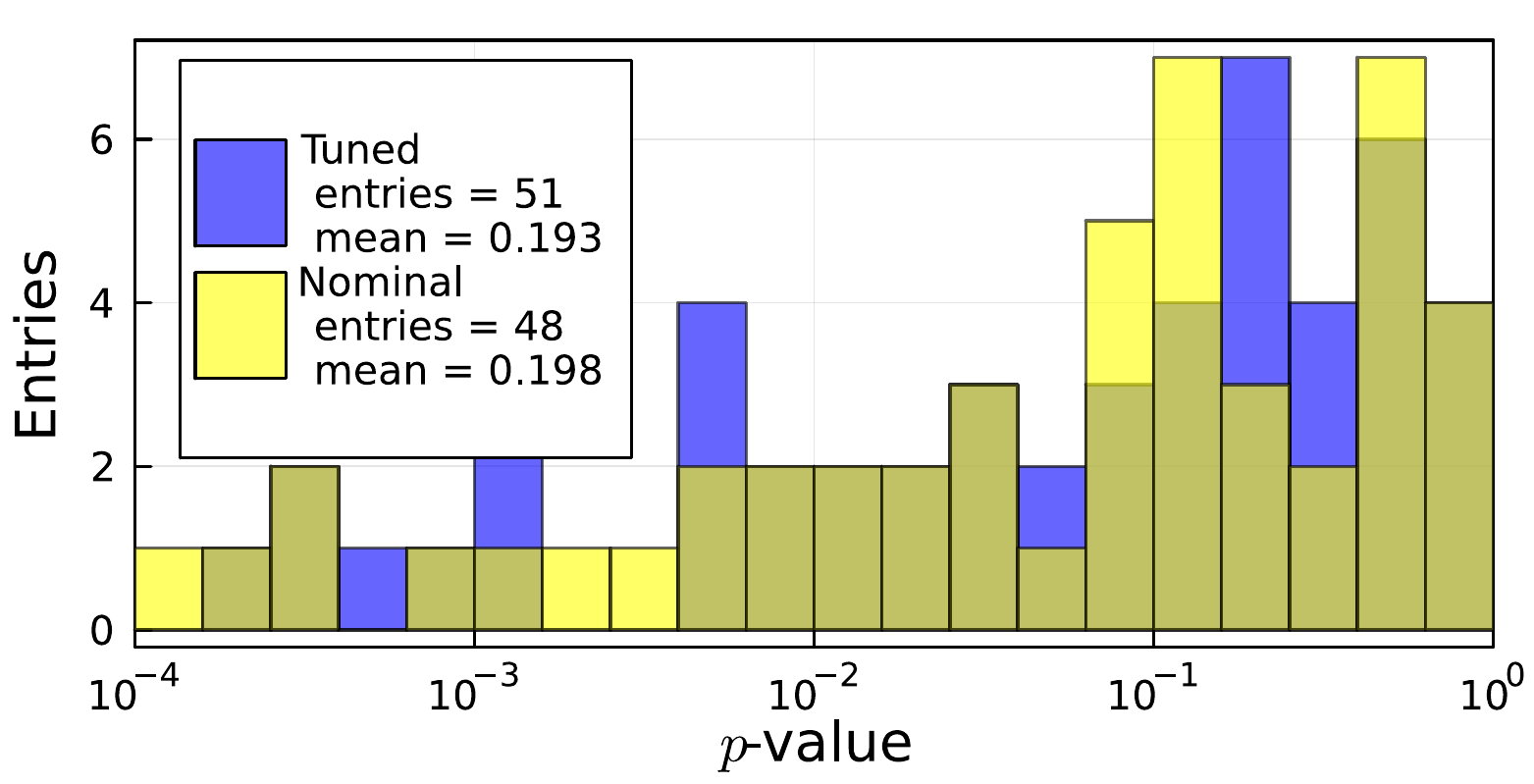}
\caption{#1}
\label{#2}
\end{figure}
}

\newcommand{\FIGthirteen}[2]{
\begin{figure}[htbp]\centering
\includegraphics[width=0.49\linewidth,height=0.28\linewidth]{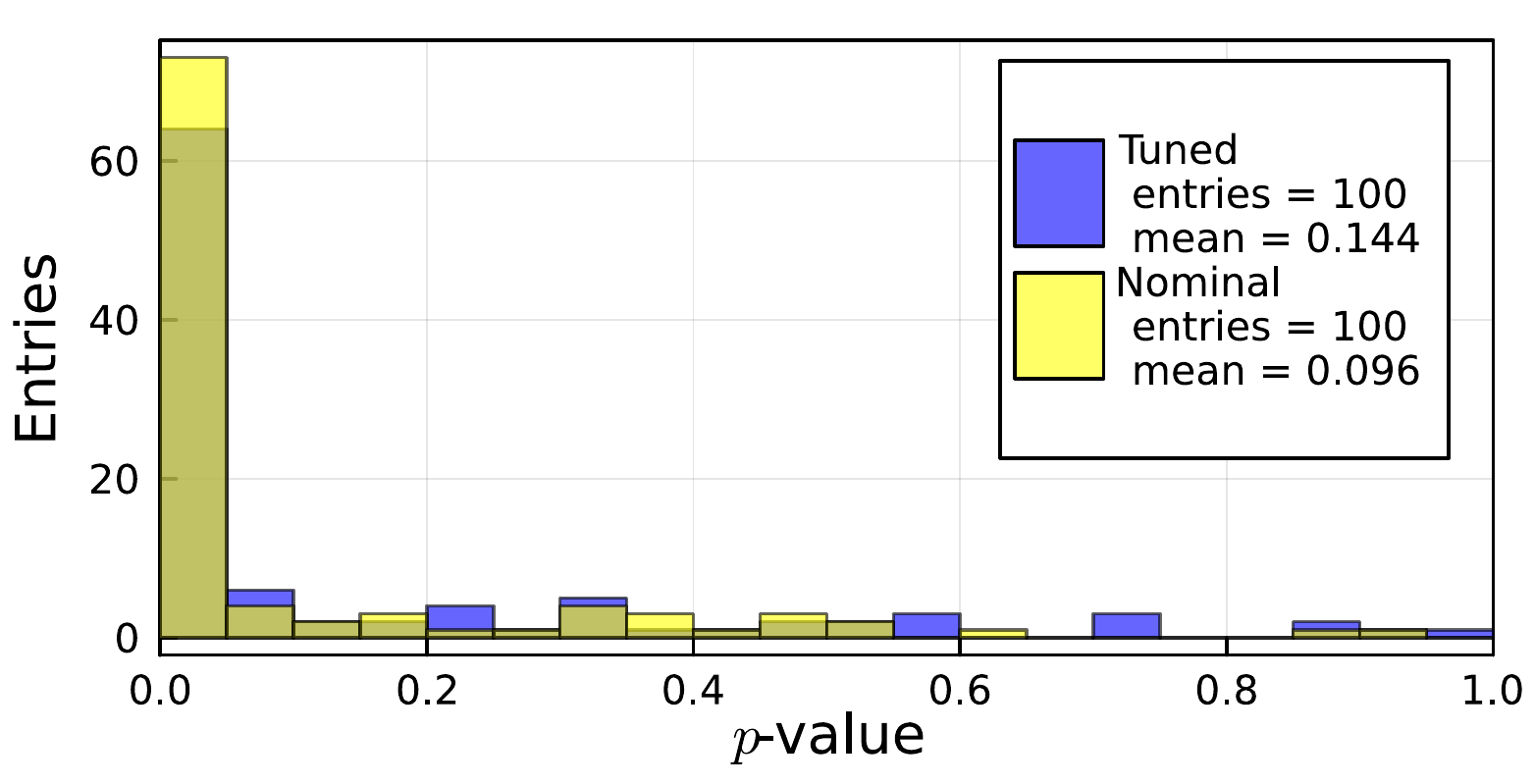}
\includegraphics[width=0.49\linewidth,height=0.28\linewidth]{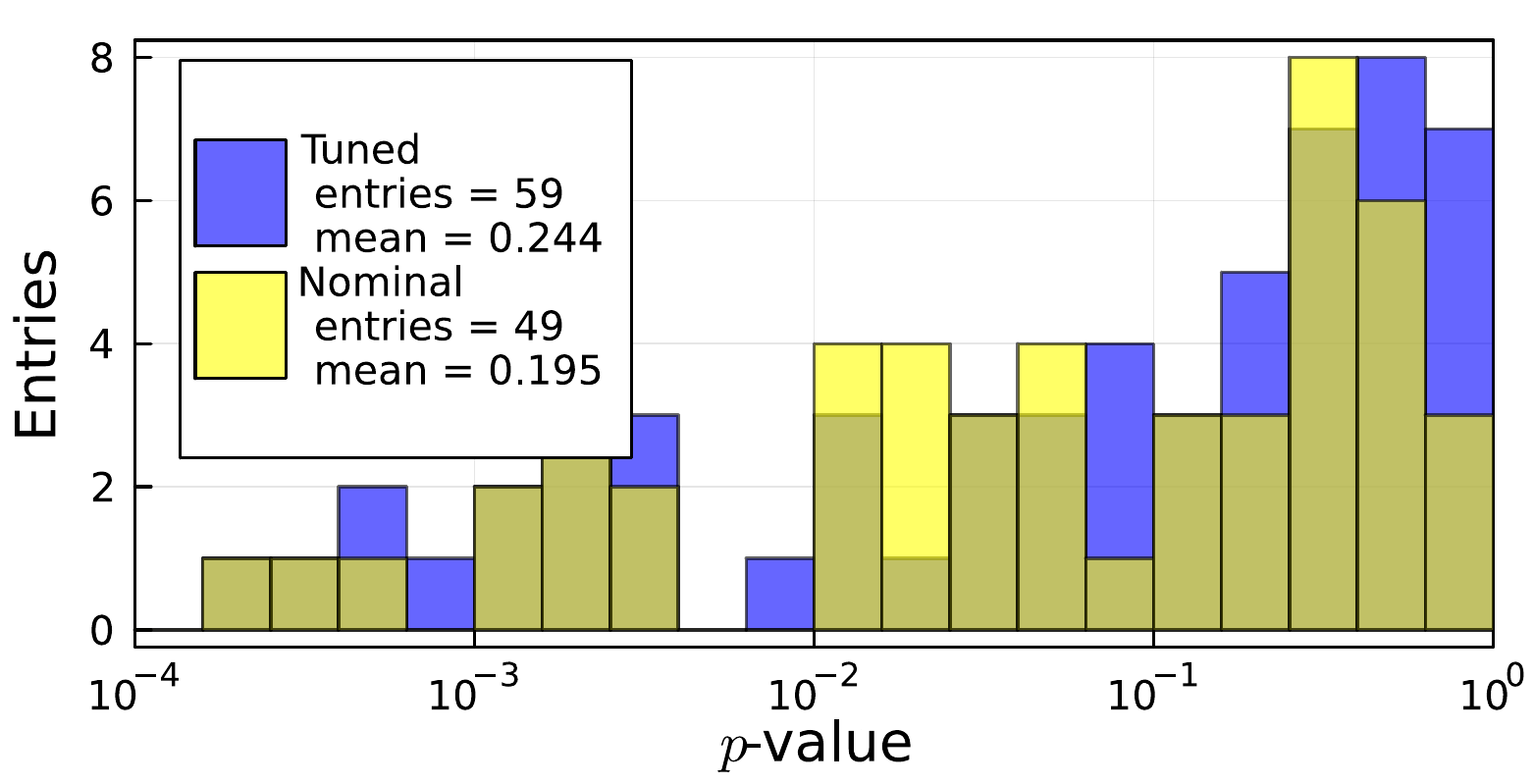}\\
\includegraphics[width=0.49\linewidth,height=0.28\linewidth]{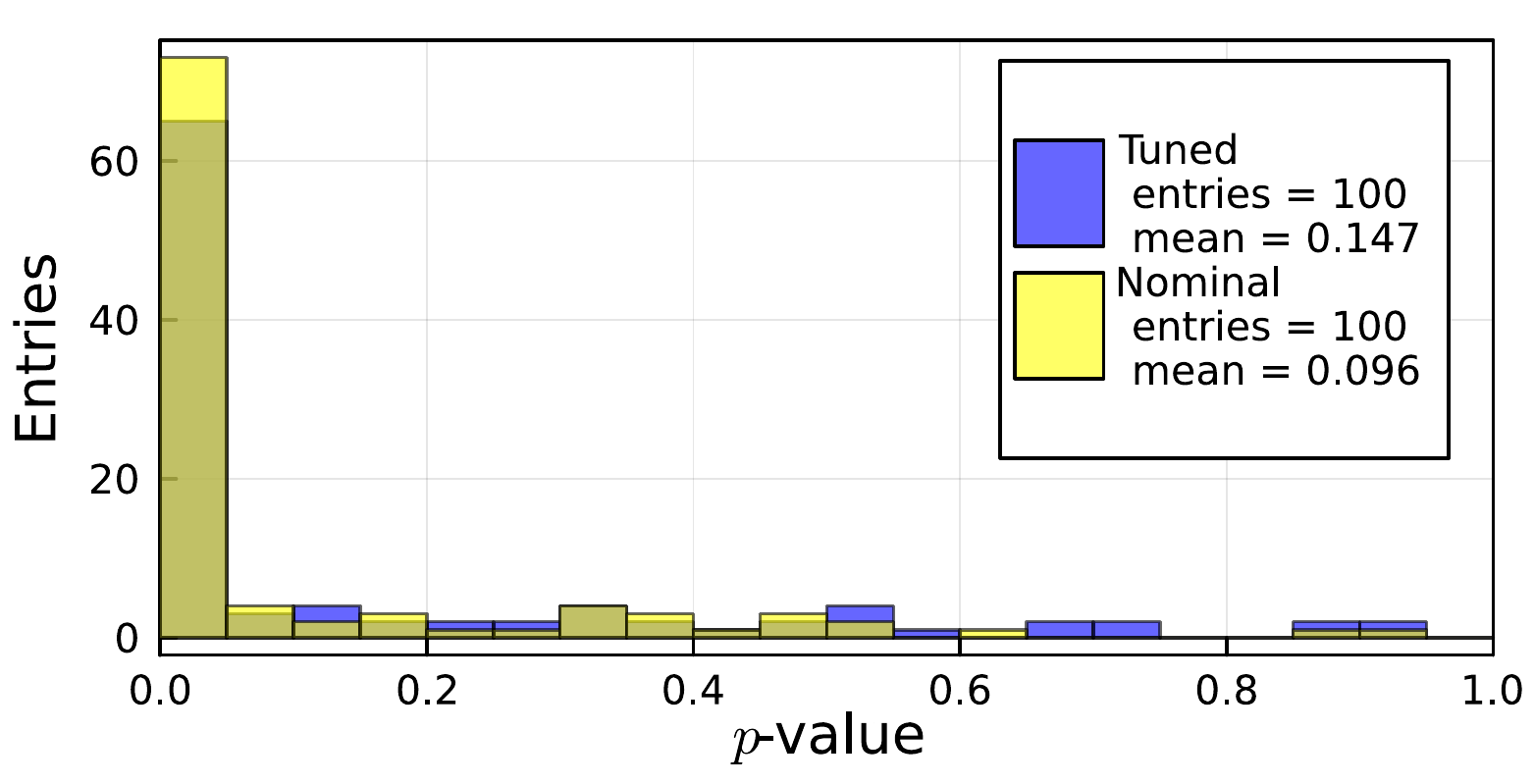}
\includegraphics[width=0.49\linewidth,height=0.28\linewidth]{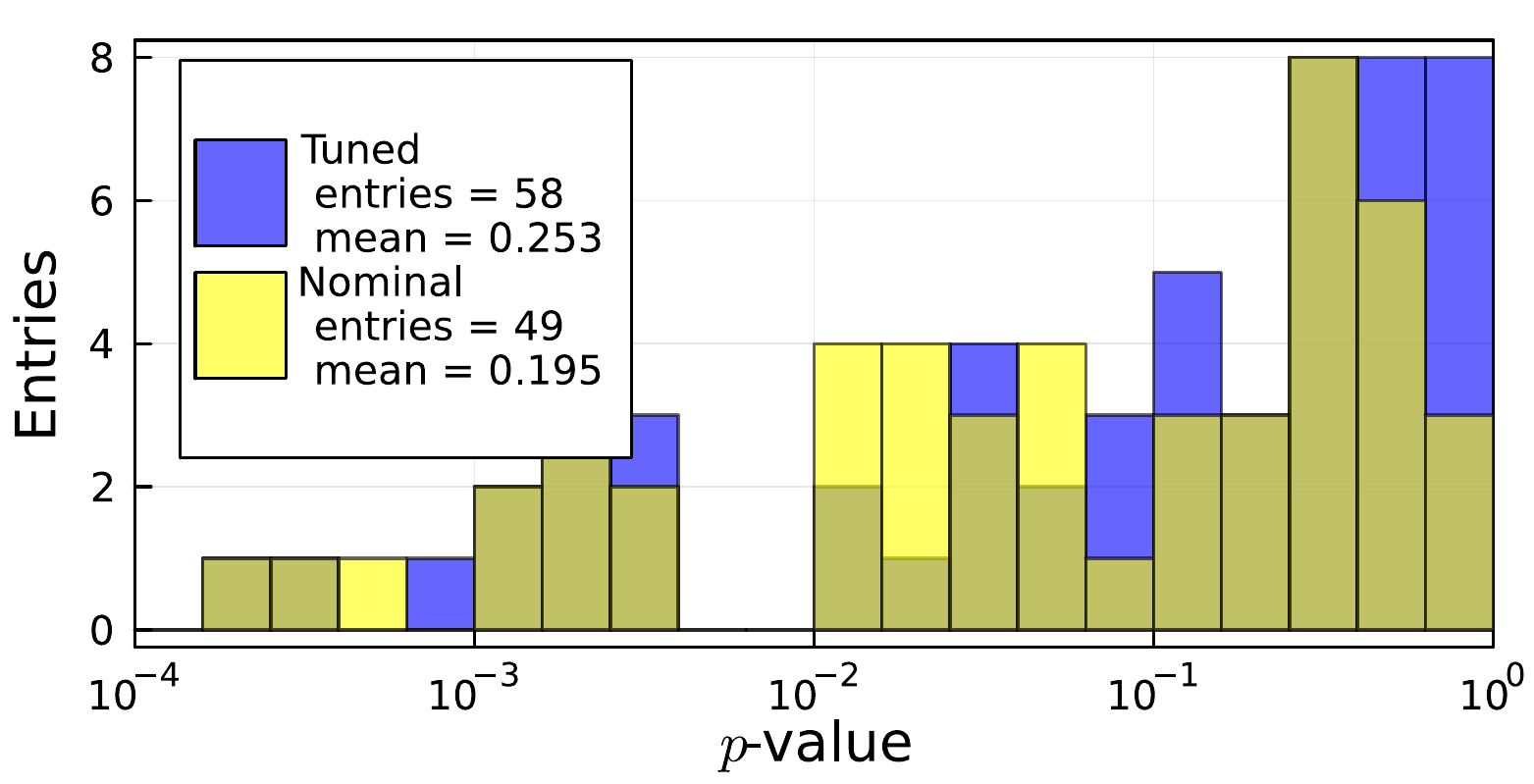}
\caption{#1}
\label{#2}
\end{figure}
}

\newcommand{\FIGfourteen}[2]{ 
\begin{figure}[htbp]\centering
\includegraphics[width=0.65\linewidth,height=0.65\linewidth]{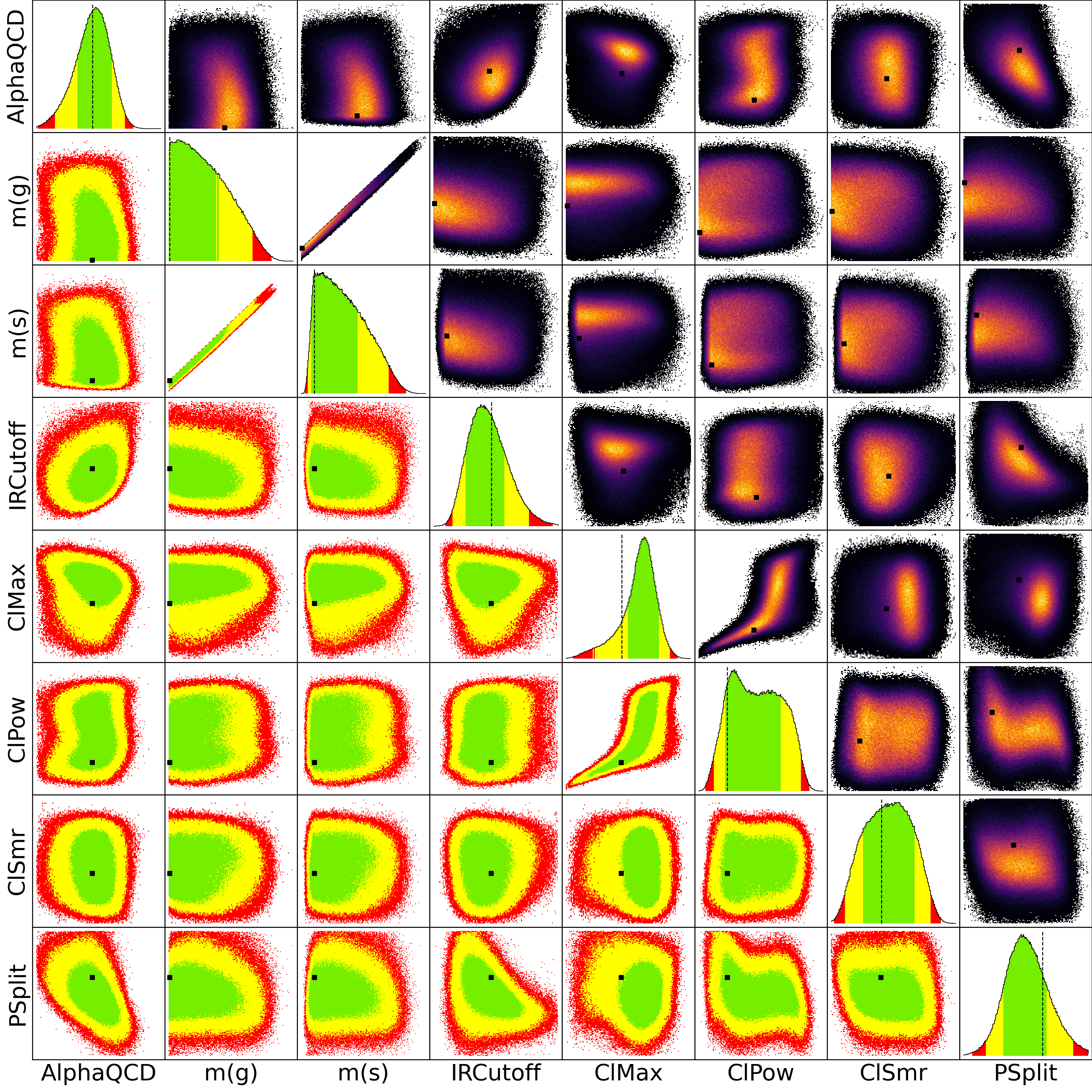}\\
\includegraphics[width=0.65\linewidth,height=0.65\linewidth]{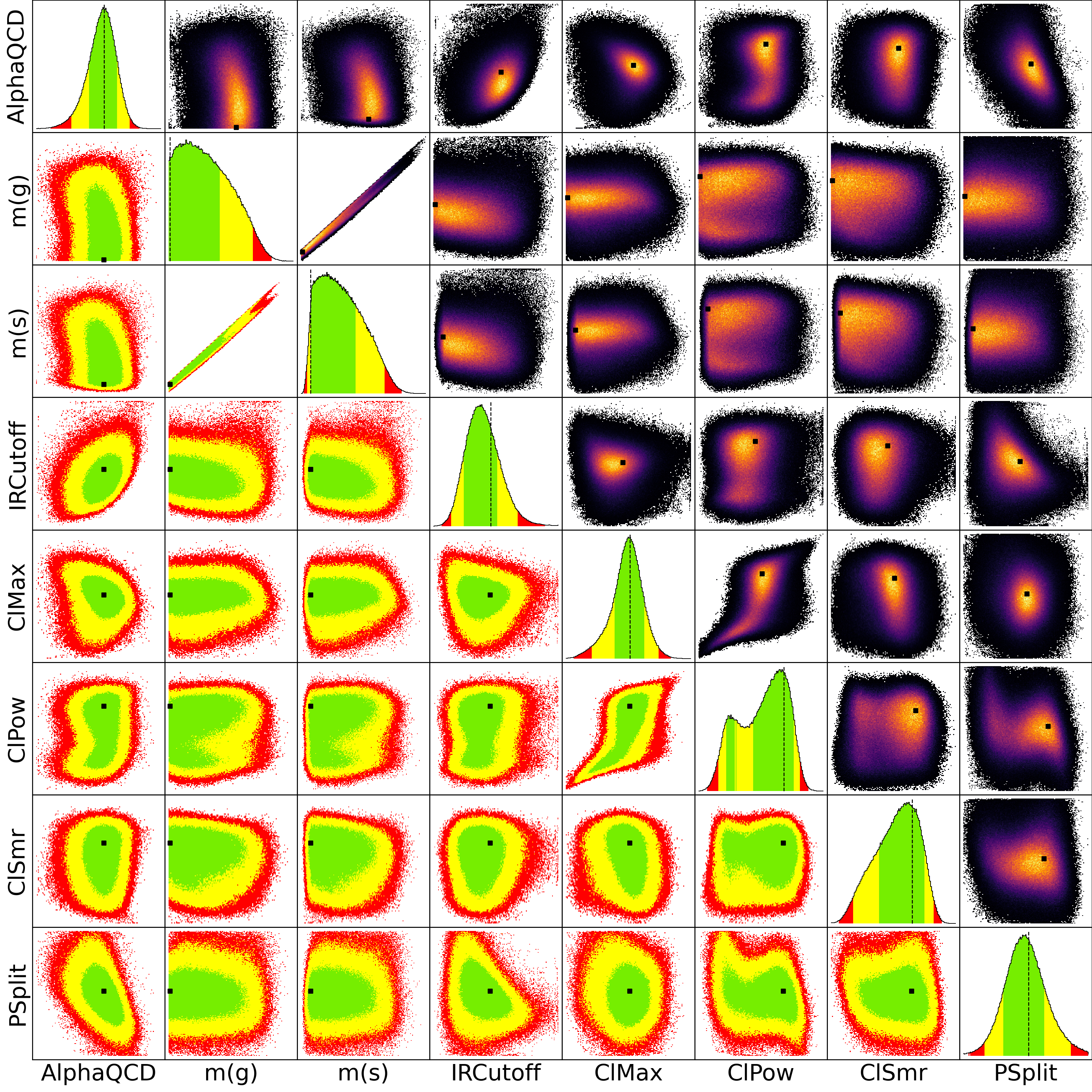}
\caption{#1} 
\label{#2}
\end{figure}
}

\newcommand{\FIGfifteen}[2]{ 
\begin{figure}[htbp]\centering
\includegraphics[width=0.65\linewidth,height=0.65\linewidth]{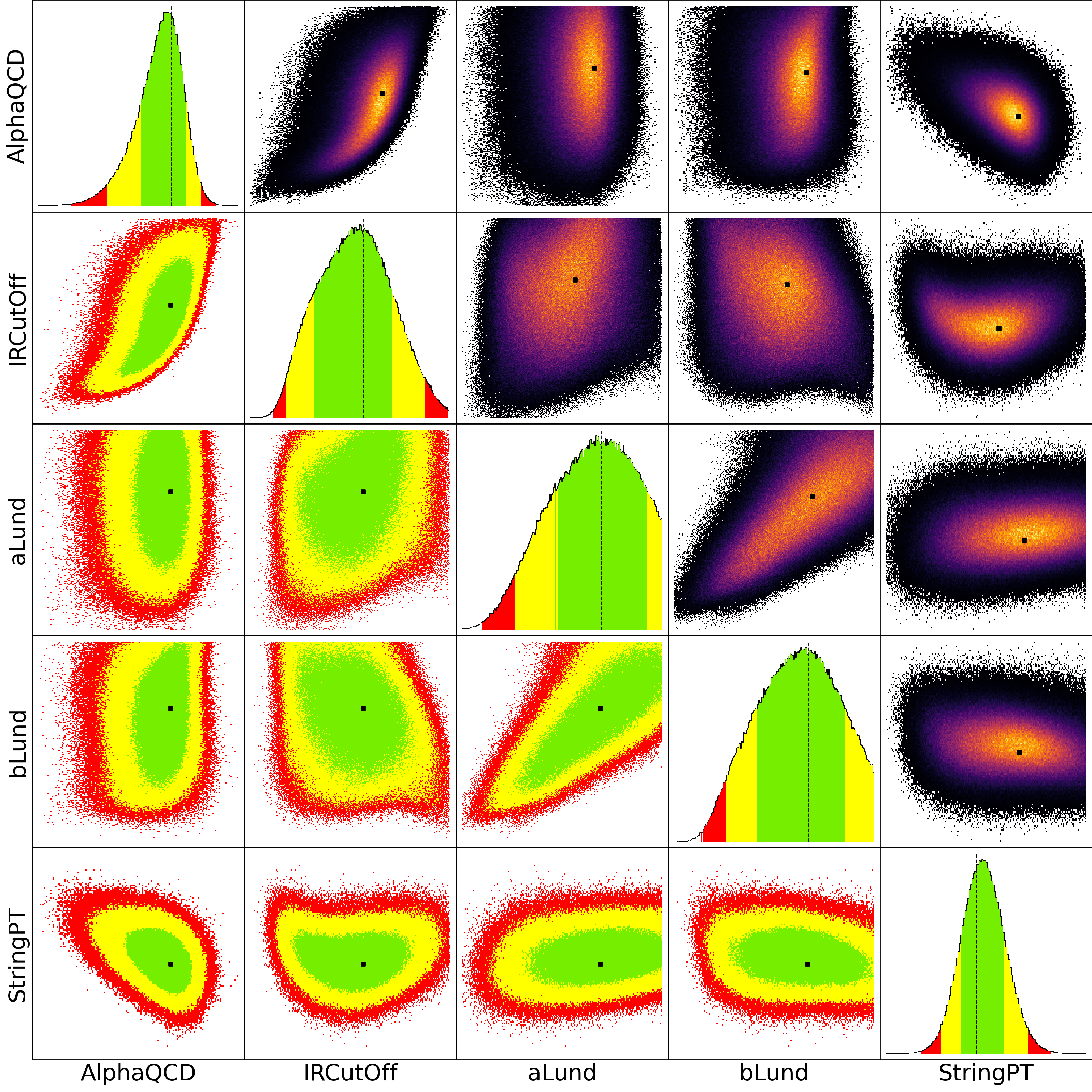}\\
\includegraphics[width=0.65\linewidth,height=0.65\linewidth]{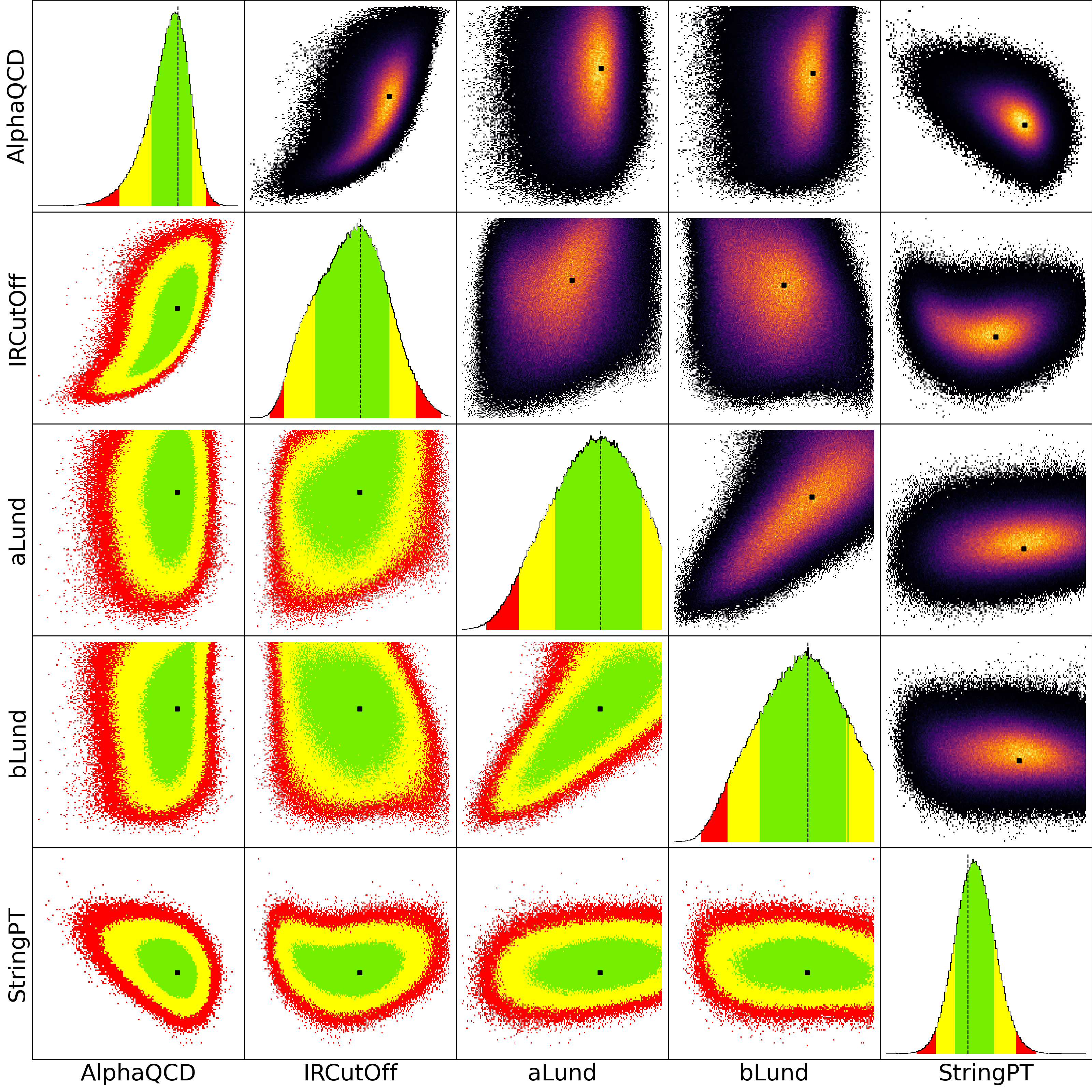}
\caption{#1} 
\label{#2}
\end{figure}
}

\newcommand{\FIGsixteen}[2]{
\begin{figure}[htbp]\centering
\includegraphics[width=0.7\linewidth]{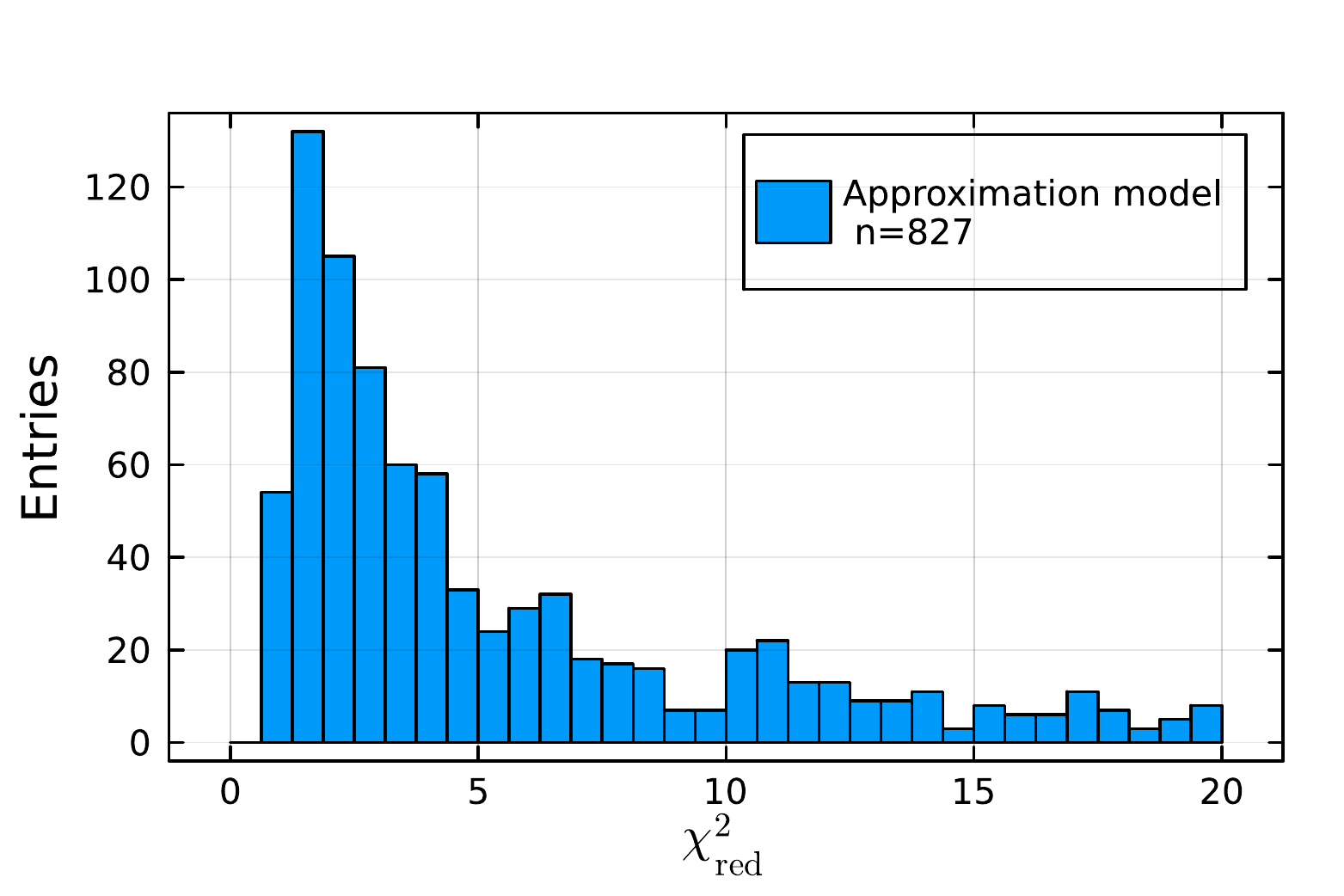}
\caption{#1}
\label{#2}
\end{figure}
}

\newcommand{\FIGseventeen}[2]{
\begin{figure}[htb]\centering
\includegraphics[width=0.7\linewidth]{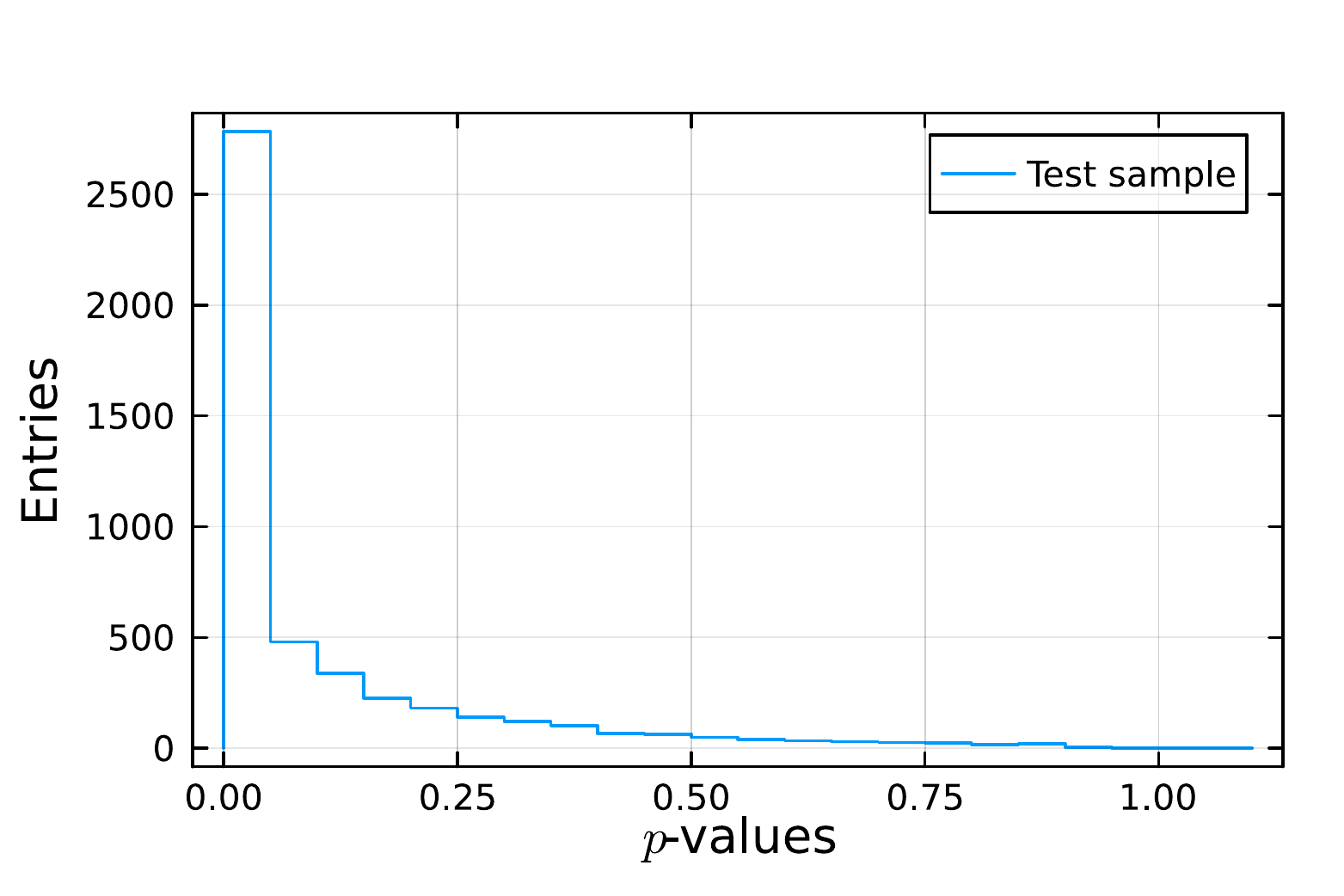}
\caption{#1}
\label{#2}
\end{figure}
}

%% file: BayesHerwigTune-def.tex
\newcommand{\eVdist}{\kern-0.06667em}



%% file: BayesHerwigTune-abs.tex
The optimisation (tuning) of the free parameters of Monte Carlo event generators by comparing their predictions with data is important since the simulations are used to calculate experimental efficiency and acceptance corrections, or provide predictions for signatures of hypothetical new processes in experiments. We present a tuning procedure that is based on Bayesian reasoning and that allows for a proper statistical interpretation of the results. The parameter space is fully explored using Markov Chain Monte Carlo. We apply the tuning procedure to the Herwig7 event generator with both the cluster and the string hadronization models and a large set of measurements from hadronic Z-boson decays produced at LEP in $e^{+}e^{-}$ collisions. Furthermore, we introduce a coherent propagation of uncertainties from the realm of parameters to the realm of observables and we show the effects of including experimental correlations of the measurements. To allow comparison with the approaches of other groups, we repeat the tuning considering weights for individual measurements.

%% file: BayesHerwigTune-txt.tex
\newpage

\section{Introduction}
\label{sec:introduction}
Monte Carlo event generators (MCEGs) are an indispensable tool in experimental and theoretical particle physics. They simulate final states in high-energy particle collisions according to the predictions of the Standard Model of Particle Physics (SM). While the interactions of partons, i.e., quarks and gluons, are calculated from first principles, the transition from partons to hadrons is modeled using a semi-empirical approach, see, e.g.\  Ref.~\cite{Buckley:2011ms} for a review. In the analysis of experimental data, the predictions from MCEGs are combined with a detailed simulation of the experimental setup, which is crucial for understanding of the data. Examples include the precise determinations of the top-quark mass and the $W$-boson mass~\cite{CDF:2022hxs,ATLAS:2018fwq}. The use of MCEGs to calculate predictions of particle physics theory to investigate a data likelihood is an application of simulation-based inference~\cite{Cranmer:2019eaq}.

Recent advances in the development of MCEGs include an improved treatment of higher-order perturbative corrections for the simulation of hard scattering processes and an improved theoretical precision of the simulation of the parton showers. To tune the simulation of parton showers, MCEGs typically have a small number of free parameters, such as the value of the strong coupling constant $\alpha_S$, the lower limit of the evolution parameter in the parton shower, and the parameters of the hadronization models. It is these parameters that have to be optimized in order to achieve the best possible description of the experimental data. 

A first complete workflow for the determination of optimal MCEG parameters was developed by the DELPHI collaboration~\cite{DELPHI:1996sen}. Today, the standard for such a workflow is a set of data-to-prediction comparison codes (analysis modules), which are included in the \texttt{Rivet} package~\cite{Bierlich:2019rhm}. The tuning workflow in the \texttt{Professor}~\cite{Buckley:2009bj} system creates multiple event samples with different parameter settings, approximates the predictions as functions of the tuning parameters for all data points, and finally finds the optimal values of the parameters by comparing the approximation functions to the data. 

The original application for tuning workflows were data from LEP including event shape observables, jet production rates, charged and identified particle multiplicities, and charged particle momentum spectra~\cite{DELPHI:1996sen,Buckley:2009bj}. In previous studies, some datasets were weighted during the final optimization step to improve the description of the data after fitting. Detailed studies of the choice of these weights were performed in Refs.~\cite{Wang:2021gdl,Krishnamoorthy:2021nwv}.  While this procedure was introduced to ensure a good description of the phenomenologically important observables, a proper statistical interpretation of the parameter estimation, in particular the uncertainty estimation, was not possible. 

The application of the \texttt{Professor} workflow on parameter subspaces was explored in Ref.~\cite{Bellm:2019owc}. Other approaches to tuning MCEGs are Bayesian optimisation~\cite{Ilten:2016csi} and machine learning-based reweighting using the complete simulated final states~\cite{Andreassen:2019nnm,Lazzarin:2020uvv}. Even though many approaches for complete tuning workflows have been proposed, manual procedures are still followed~\cite{Skands:2014pea}. 

In this study, a tuning procedure based on Bayesian reasoning that allows for a proper statistical interpretation of the results was developed. Using Markov chain Monte Carlo algorithms, the full posterior distributions for the MCEG parameters were determined. Those posterior distributions provide a better understanding of the parameter space and may also reveal the limitations of the underlying physical models. In addition, the full posterior distribution allows uncertainties to be coherently propagated from the parameter space to the space of observables. As a result, the tuning procedure provides not only an estimate of the optimal MCEG parameters (with uncertainties and correlations), but also uncertainty estimates for the predictions. As a study case, the developed tuning procedure  was applied to a well-defined set of measurements and the parameters of the \texttt{Herwig7} MCEG with two different hadronization models were optimized. In addition, the effects of including experimental correlations of the measurements were studied. To allow comparison with the approaches of other groups, the tuning was repeated considering the weights for individual measurements.

\section{Data selection}
Similar to previous tuning studies~\cite{Fischer:2014bja,Klimpel:2018ydj,Buckley:2009bj}, the data measured in $e^{+}e^{-}$ collisions were used. The main motivation for this choice was to avoid ambiguities in the choice of parton density functions, which are unavoidable when using the data from proton-proton or proton-lepton colliders. In addition, the process $e^{+}e^{-}\rightarrow (Z/\gamma)^* \rightarrow {\rm partons}$ is very well understood in perturbative QCD (pQCD), and in modern MCEGs it is used together with parton showers for accurate and stable simulations of the process $e^{+}e^{-}\rightarrow (Z/\gamma)^* \rightarrow {\rm hadrons}$.

The focus of the study is on the tuning of hadronization models used in the MCEGs. Therefore, a set of measurements that can be used to assess the quality of the modeling of hadronic final states~\cite{Bierlich:2019rhm,ALEPH:1996oqp,ALEPH:2001pfo,DELPHI:1996sen,JADE:1999zar,ParticleDataGroup:2008zun} was selected. The  corresponding \texttt{Rivet} analysis modules are
ALEPH\_1996\_S3486095~\cite{ALEPH:1996oqp}, 
ALEPH\_2001\_S4656318~\cite{ALEPH:2001pfo},
DELPHI\_1996\_S3430090~\cite{DELPHI:1996sen},
JADE\_OPAL\_2000\_S4300807~\cite{JADE:1999zar} and 
PDG\_HADRON\_MULTIPLICITIES~\cite{ParticleDataGroup:2008zun}.
The \texttt{Rivet} analysis modules ALEPH\_1996\_S3486095 and 
DELPHI\_1996\_S3430090 provide calculations for the  event shape observables such as sphericity, thrust, aplanarity etc. These observables and the differential jet rates provided by the JADE\_OPAL\_2000\_S4300807 analysis module are  sensitive to $\alpha_S$. The ALEPH\_1996\_S3486095 analysis module also provides identified particle spectra  sensitive to fragmentation parameters. The $b$-quark fragmentation function observables were calculated with the ALEPH\_2001\_S4656318 analysis module. 
The PDG\_HADRON\_MULTIPLICITIES module was used to study the simulation of particle multiplicities.

In total, $100$ observables from these five \texttt{Rivet} analysis modules were used. 
The full list of the used observables is given in Tabs.~\ref{table:longlistone},~\ref{table:longlistthree} and~\ref{table:longlisttwo} in Appendix~\ref{sec:apx_observables}.
  
\section{Monte Carlo event generators}
\label{sec:MonteCarlogeneration}
To perform the calculations for the $e^{+}e^{-}\rightarrow (Z/\gamma^*) \rightarrow {\rm 2,3,4,5\ partons}$ processes the \texttt{Herwig7} MCEG version 7.2.2~\cite{Bellm:2015jjp} was used with the MENLOPS method~\cite{Hamilton:2010wh} using the \texttt{MadGraph5}~\cite{Alwall:2011uj} matrix element generator and the \texttt{OpenLoops}~\cite{Cascioli:2011va} one loop library. The two-parton final states were predicted with full NLO accuracy in perturbative QCD in this scheme. The QCD matrix elements were calculated with mass effects taken into account for massive $b$-quarks. Two models were used for the modeling of the hadronization process and the details on these models are given below. The simulated events produced by the MCEG were put into the \texttt{HepMC} format~\cite{Buckley:2019xhk} and passed to the \texttt{Rivet} package, where they were processed by the \texttt{Rivet} analysis modules for the corresponding data sets. The comparison between the MCEG predictions and the data is explained in Section~\ref{sec:Statistical framework}. 

\subsection{\texttt{Herwig7} with the cluster hadronization model}
The default hadronization model of the \texttt{Herwig7} MCEG is the cluster hadronisation model~\cite{Webber:1983if}. This model is referred to as \texttt{Herwig7-H7}. 
For this model, the list of parameters, as well as the ranges in which the parameters were varied, are listed in Tab.~\ref{table:parameterH7}. 

\TABparamsher

The selection of the parameter ranges was driven by two factors: the desire for a wide coverage of the 
physically meaningful parameter space and the constraints imposed on the parameters by the MC generator code and models. As a result, most parameter ranges were chosen to be in a $\pm 50\%$ window around their default settings 
with some exceptions that will be discussed in the following.
The ranges for the \texttt{AlphaQCD} parameter were chosen taking into account the hardcoded minimal and maximal values in the generator code. 
Similarly, the gluon and strange quark constituent masses are limited
by the boundaries
set by the hadronization model~\cite{Bahr:2008pv}.
The implementation of the model in \texttt{Herwig7} also requires the gluon and the strange quark to fulfill the condition $m_{g} > \frac{m_{s}}{2}$ for a successful run.
Hereby, a lower bound can be set with the condition $m_{s} > m_{u,d} = 0.35$.
Given the constraints of the constituent masses, $m_{s}$ was chosen to be varied as a fraction of $m_{g}$ with 
the lower fraction equal to the default ratio of $m_{s}/m_{g} = 0.47$ which gives a lower bound of $\min(m_{g}) = 0.74$.

\subsection{\texttt{Herwig7} with the Lund string hadronization model.}
The second hadronization model available for \texttt{Herwig7} is the 
Lund string hadronization model as implemented in \texttt{Pythia8}~\cite{Bierlich:2022pfr}.
The \texttt{Pythia8} hadronization code is interfaced to \texttt{Herwig7} using the \texttt{TheP8I} interface~\cite{TheP8I,Bellm:2019owc}. This model is referred to as \texttt{Herwig7-P8} and used \texttt{Pythia~8.306} to generate the \texttt{Herwig7-P8} samples. 
The selection of the parameter ranges for the \texttt{Herwig7-P8} model was done in the same way as for the \texttt{Herwig7-H7} model.
The ranges for the tune were chosen within the allowed scope of the \texttt{Pythia8} framework. The parameters \texttt{aExtraDiQuark} and \texttt{aExtraSQuark} have been varied and fitted as well. However, due to a lack of sensitivity of the observables to these parameters, they were  fixed when running the tune. Their fixed value was derived from the Monash Tune~\cite{Skands:2014pea} with $0.0$ for \texttt{aExtraSQuark} and $0.97$ for \texttt{aExtraDiQuark}. The parameter ranges for \texttt{Herwig7-P8} are shown in Tab.~\ref{table:parameterP8}. 

\TABparamspytFix

\section{Tuning procedure and statistical model} 
\label{sec:Statistical framework}
The tuning procedure was implemented in the \texttt{JULIA} package \texttt{BAT.jl}~\cite{Schulz:2021BAT} which is a software tool for Bayesian analysis containing algorithms for parameter estimation, hypothesis testing, model comparison, and goodness-of-fit tests. \texttt{BAT.jl} provides interfaces to define arbitrary data likelihood functions and prior distributions for statistical models defined by the user. The data likelihood and the prior were multiplied to obtain the (unnormalized) posterior distribution which was then explored with dedicated algorithms, in particular Markov Chain Monte Carlo techniques and derivatives thereof.

The \texttt{BAT.jl}-based package \texttt{EFTFitter.jl}~\cite{EFTfitter2016} provides a data likelihood function for combining several quantities including uncertainties and correlated data. It has been developed for the interpretation of data in the context of effective field theories, which is a mathematically similar problem. The data likelihood $L(\vec{D}|\vec{\lambda})$ is a multivariate Gaussian, i.e.\ 
\begin{equation}
  \ln L(\vec{D}|\vec{\lambda}) = - \frac{1}{2} [\vec{D} - \vec{f}(\vec{\lambda})]^T \cdot M^{-1} \cdot
  [\vec{D} - \vec{f}(\vec{\lambda})] \; .
  \label{eqn:eftfitterl}
\end{equation}
The data are represented by $\vec{D}$, where each component of the vector corresponds to a measured value of an observable (either a single measurement or a bin of a differential distribution). The covariance matrix of the data is denoted by $M$. The components ${f}_{b,O}(\vec{\lambda})$ of the vector $\vec{f}$ represent the MCEG predictions of bin $b$ for observable $O$ as a function of the parameters $\vec{\lambda}$. For the sake of simplicity and to facilitate comparison with previous results, the prior distributions of the MCEG parameters are chosen to be uniform over the ranges shown in Tabs.~\ref{table:parameterH7} and~\ref{table:parameterP8}. For the \texttt{Herwig7-H7} tune, an additional constraint is imposed on the data likelihood by demanding $m_{g} > m_{s}/2$ in the generated samples. The covariance matrix was chosen to be a diagonal matrix, the case of non-negligible off-diagonal elements is discussed in Section~\ref{sec:correlation}.

The posterior distribution was sampled using the Metropolis-Hasting (MH) algorithm for which convergence is achieved within a few tuning cycles. The MH sampling was performed using six chains with $10^6$ sampling steps. In order to check for convergence of the different chains, \texttt{BAT.jl} uses the Gelman-Rubin test~\cite{Gelman:1992zz}, which has been generalized for the multivariate case by Brooks and Gelman~\cite{Brooks:1998}.
The convergence parameter $R$, which is calculated by comparing the variance of the samples within a chain to the variance of samples between different chains, see Ref.~\cite{Brooks:1998}, was slightly increased to $1.3$ from its default value of $1.1$ to account for the larger number of chains which leads to higher distances, causing higher $R$ values of the chains during tuning. 

The global mode values of the posterior distribution were chosen as the optimal parameter set. While the full multidimensional posterior distribution can be used for further investigations, e.g.\  studies of global modes or the propagation of uncertainties, the uncertainties for the individual parameters were defined as the smallest $68$\% credibility interval of the marginalized distributions. 

\section{Parameterization of the MCEG predictions}
\label{sec:Approximation}

The data likelihood contains a vector of functions $\vec{f}$ that represents the MCEG predictions as a function of the free parameters, e.g.\  the mean charged multiplicity as a function of $\alpha_{s}$. 
Since the evaluation of these predictions using MC simulated events is very CPU-time intensive, they were replaced by approximative parametrizations. The resulting analytic expressions for the functions $f_{b,O}$ are much faster to calculate during the sampling of the parameter space during the optimization process.

To parameterize the MCEG response the multidimensional cubic polynomials were used, i.e. functions of the form
\begin{align}
  f_{b,O}(\vec{\lambda}) \approx f_{b,O}^{\mathrm{cubic}}(\vec{\lambda}) = c_0 &+ \sum_i c_i \lambda_i + \sum_i \sum_{j \leq i} c_{ij} \lambda_i \lambda_j 
  + \sum_i \sum_{j \leq i} \sum_{k \leq j} c_{ijk} \lambda_i \lambda_j \lambda_k
  \label{eq:fit}
\end{align}
for each bin $b$ and observable $O$. The parameters $\vec{\lambda}$ for the used hadronization models are shown in Tabs.~\ref{table:parameterH7} and  ~\ref{table:parameterP8}. The linear, quadratic and cubic polynomial coefficients are denoted $c_i$, $c_{ij}$ and $c_{ijk}$, respectively. The total number of coefficients needed for the  approximation with $N$ parameters was 
$1+N+N(N+1)/2+N(N+1)(N+2)/6$. 

As the set of reference points for the fit, $500$ ($700$) randomly chosen parameter sets were used for the parametrization of the \texttt{Herwig7-H7} (\texttt{Herwig7-P8}) models. For each set $10^{6}$ MC events were generated. Below, those samples are referred to as {\it analysis samples}. For each resulting sample of events, the predictions for each observable were calculated using the \texttt{Rivet} framework. Finally, the polynomial model was fitted to the MCEG predictions using the \texttt{LsqFit.jl}~\cite{LsqFit} package, which implements the Levenberg-Marquardt algorithm~\cite{Marquardt:1963,Levenberg:1944} for non-linear fitting procedures.

A reasonable agreement between the MCEG predictions and the corresponding parametrization functions was observed. Studies of the goodness-of-fit can be found in Appendix~\ref{sec:apx_ipol}. While the focus of the current study is on the tuning process itself, it is worth noting that for a high-precision tune that is used, e.g., by experimental collaborations, the parametrization of the predictions should be addressed in more detail and alternative techniques, e.g.\ template morphing, should be considered.

To give a visual impression of the agreement between the MCEG predictions and the parametrization functions, additional {\it test samples} were produced for which only one parameter at a time was varied. All parameters were set to their default values according to Tabs.~\ref{table:parameterH7} and~\ref{table:parameterP8}, and one parameter was varied in the range given there in eleven equidistant steps. The first bin content of a sphericity distribution as a function of four parameters is shown  in Fig.~\ref{fig:gridfit} as an example. The markers indicate the MC calculation and the red line represents the fitted parameterization function. The red area represents the uncertainties from the fitting procedure. To guide the eye, the blue area represents the range of predicted values from the parametrization model when changing the default values by $\pm 5$\%. In general, the parameterization functions describe the MCEG predictions reasonably well with deviations of the order of a few percent.

\FIGone{The content of the first bin of the sphericity distribution from the \texttt{Rivet} module DELPHI\_1996\_S3430090 as a function of \texttt{AlphaQCD}, \texttt{ClSmr}, \texttt{ClMax} and \texttt{ClPow} with parameter sets from the test samples. The approximation model is shown in red in comparison to the MCEG test sample predictions. The red band represents the uncertainty obtained by propagating the uncertainty of the fitted coefficients. The blue band represents variations caused by shifting the default values for the evaluation of the approximation model by $\pm 5$\%.}{fig:gridfit}

\section{Propagation of uncertainties}
\label{sec:uncertainties}
One of the main advantages of using a Bayesian approach in combination with a fast parametrization of the predictions is the opportunity to propagate the uncertainty from the parameter space to the space of observables. The resulting uncertainty is then related to the tuning procedure and limited knowledge about the MCEG parameters. The uncertainty propagation is done by re-sampling the posterior distribution to generate $10^6$ parameter points $\vec{p}$. 
For each bin of each observable, these points were evaluated according to the parametrization function with the corresponding coefficients $\vec{y}_{b,O} = f_{b,O}(\vec{p})$ using Eq.~\eqref{eq:fit}.
The resulting values $\vec{y}_{b,O}$ represent the statistical distribution of the bin content and the width of this distribution reflects the statistical uncertainty from the tuning process.

As an example, Fig.~\ref{fig:errorprop} shows these distributions for the first bin of the sphericity observable (top) and for the multiplicity of $B^+_u$ mesons (bottom). One often uses the standard deviation as a measure of uncertainty, silently assuming a Gaussian distribution.
In the case of the sphericity observable, this assumption is roughly valid as the distribution is uni-modal and symmetric. However, in the case of the $B^+_u$-meson multiplicity, the distribution has two separate modes and is not symmetric. In such cases, a full propagation of the uncertainty is required even if it is difficult to display graphically. For simplicity, the standard deviation was used as a measure of uncertainty for the tuning process in the following.  

\FIGtwo{Distribution of the bin content for two bins of two observables using the parametrization with the posterior samples as input. The figure on top shows the first bin of the sphericity observable while the bottom figure shows the multiplicity of $B^+_u$ mesons.}{fig:errorprop}

\section{Results}
\label{sec:results}

We perform the tune of the \texttt{Herwig7-H7} and the \texttt{Herwig7-P8} models according to the procedure defined above. The global modes are quoted as the final results of the tune, while the modes and smallest 68\% intervals of the marginalized distributions are quoted as results for the individual parameters and their uncertainties. We also display and discuss the one- and two-dimensional marginalized distributions. In order to evaluate the quality of the tune, we generate a new MC sample with the parameters set to the global mode and compare each data and MCEG distribution by calculating the $\chi^{2}$ and the corresponding $p$-value.

\subsection{\texttt{Herwig7-H7} tune}
\label{sec:H7results}

The global mode as well as the marginal and the smallest 68\% credibility intervals of the marginalized posterior probability are summarized in Tab.~\ref{table:resultTuneHer}. The global and marginalized modes are close. The marginalized mode values are in agreement with the default settings shown in Tab.~\ref{table:parameterH7} within the uncertainties for each parameter except for \texttt{ClSmr}. For that parameter, the default setting of $0.3437$ lies within the $95$ percentile.

\TABresultTuneHer 

The one- and two-dimensional marginalized distributions are shown in Fig.~\ref{fig:resultsHer}. None of the parameters are described by a normal distribution. The distributions for the parameters \texttt{AlphaQCD}, \texttt{IRCutoff} and \texttt{PSplit} have only a slight asymmetry in their distributions, while \texttt{ClSmr} and \texttt{ClMax} show a more pronounced asymmetry. The distributions representing the gluon and strange quark constituent masses, $(m_{g}$ and $m_{s})$, are almost one-sided with a sharp increase and slow drop-off towards higher values. The marginalized distribution for \texttt{ClPow} is multimodal. The global mode is located in the first peak, while the default value is closer to the second peak. Most two-dimensional distributions show a mild linear correlation. The most prominent exceptions are a very strong correlation between $m(g)$ and $m(s)$ and the highly non-Gaussian shape of the distributions of \texttt{ClPow} and \texttt{ClMax}.

\FIGthree{One and two-dimensional marginalized posterior distributions of the parameters for the  tune of the \texttt{Herwig7-H7} model. The green, yellow and red areas contain the smallest $68$, $95$ and $99$\% intervals of the marginalized probability distributions, respectively. The dots and the lines are projections of the global mode representing the point with the highest probability.}{fig:resultsHer}

The distribution of the $p$-values for each observable using the \texttt{Herwig7-H7} model with tuned and nominal parameter sets are shown in Fig.~\ref{fig:resultsHerProb}. The mean $p$-value increases from $0.095$ for the default parameter values to about $0.133$ for the tuned values. Since a majority of observables still tend to have low $p$-values we compare the two parameter sets on a logarithmic scale.

\FIGfour{Distributions of $p$-values for the tuned and nominal MC samples for the \texttt{Herwig7-H7} tune. Each observable contributes one $p$-value toward the histogram. The bottom figure shows a subrange of the distribution $p > 10^{-4}$ in a logarithmic scale.}{fig:resultsHerProb}

To demonstrate the impact of the tuning, Fig.~\ref{fig:restunedobs} shows two observables as an example, namely the sphericity (left) and the $B^+_u$ multiplicity (right), calculated from the tuned and the nominal MCEG sample as well as from the data. The uncertainty bands for the tuned sample contain the MC statistical uncertainty and the propagated uncertainty from the tuning process as discussed in Section~\ref{sec:uncertainties}. The uncertainties associated with the tuned sample are thus larger than those associated with the nominal sample. The tuned MCEG model results either in the same or in a better agreement with the data compared to the results obtained with the nominal sample. 
We find values of $\chi^2 / ndf$ after (before) tuning of 7.60 (9.82) for the sphericity distribution and of 3.86 (25.6) for the mean $B^+_u$ multiplicity. 
This trend is seen in most observables and expected from the average $p$-value distribution. 

\FIGfive{Distribution of the sphericity observable from DELPHI~\protect\cite{DELPHI:1996sen} (left) and the mean $B^+_u$ multiplicity~\protect\cite{ParticleDataGroup:2008zun} (right) for the data and the tuned and nominal MCEG samples for the \texttt{Herwig7-H7} tune. The calculation of the uncertainties for the nominal and tuned results is explained in the text. The bottom sections of the figures show the ratio to data.}{fig:restunedobs}

\FloatBarrier

\subsection{\texttt{Herwig7-P8} tune}
\label{sec:P8results}

The marginal mode and smallest $68$\% intervals for the \texttt{Herwig7-P8} tune can be found in Tab.~\ref{table:resultTunePyt}. For \texttt{AlphaQCD}, \texttt{SigmaPT}, \texttt{IRCutoff} and \texttt{aLund} the default values, listed in Tab.~\ref{table:parameterP8}, are within the smallest $68$\% intervals while the \texttt{bLund} default values are significantly smaller.

\TABresultTunePytFix

Fig.~\ref{fig:resultsPyt} shows the posterior distributions for the \texttt{Herwig7-P8} tune.
The \texttt{AlphaQCD} and \texttt{SigmaPT} parameters are well constrained with both marginalized distributions being almost symmetric and showing a smaller width compared to the full prior size. The \texttt{IRCutoff} variable shows similar behavior. \texttt{aLund} and \texttt{bLund} show a correlation as visible in their two-dimensional marginalized distribution.
Contrary to the other variables, their constraint is rather weak as the one-dimensional marginal distribution has a larger width which results in them being cut-off by the prior edges towards higher values.
Compared to the posterior of the \texttt{Herwig7-H7} tune in Fig.~\ref{fig:resultsHer}, the posterior has a single mode indicating a less ambiguous solution tune. 

\FIGsix{Results for the tune of the \texttt{Herwig7-P8} hadronization model. See the caption on Fig.\ref{fig:resultsHer} for a detailed explanation. 
}{fig:resultsPyt}

The distributions of the $p$-values for the \texttt{Herwig7-P8} tune compared to the default parameter set are shown in Fig.~\ref{fig:resultsP8IProb}. Similarly to the \texttt{Herwig7-H7} tune, the mean of the $p$-values increases from $0.096$ to $0.143$ for the tuned MCEG with fewer observables showing a $p$-value below $p < 10^{-4}$.
As examples, the sphericity and $B^+_u$ multiplicity observables for the tuned and nominal samples are shown in Fig.~\ref{fig:resdistPyt}.
In this case, the sphericity distribution shows an improved agreement to data, while the multiplicity distribution shows no improvement. 
We find values of $\chi^2 / ndf$ after (before) tuning of 6.42 (16.6) for the sphericity distribution and of 13.5 (13.3) for the mean $B^+_u$ multiplicity.
However, the overall agreement to the data improves as indicated by the $p$-values.

\FIGseven{Distributions of $p$-values for the tuned and nominal MC samples for the \texttt{Herwig7-P8} tune. Each observable contributes one $p$-value toward the histogram. The right figure shows a subrange $p > 10^{-4}$ of the distribution in a logarithmic scale.}{fig:resultsP8IProb}

\FIGeight{Distribution of the sphericity observable from DELPHI~\protect\cite{DELPHI:1996sen} on the left and the mean $B^+_u$ multiplicity~\protect\cite{ParticleDataGroup:2008zun} on the right for the data and the tuned and nominal MCEG samples for the \texttt{Herwig7-P8} tune. See the caption on Fig.~\ref{fig:restunedobs} for further details.}{fig:resdistPyt}

\subsection{\texttt{Herwig7-H7} and \texttt{Herwig7-P8} comparison}
\label{sec:compresults}

In addition to the comparison of the tuned and nominal samples, both hadronization models can be compared to each other. While it is not possible to compare the parameters one to one, the distributions of the observables and their agreement with the data can be compared.
The distributions of the $p$-values of the observables suggest that the \texttt{Herwig7-P8} model has a better overall agreement to data, as can be seen in Fig.~\ref{fig:resultsprobComp}, with an increase in $p$-value of about $8\%$ when compared to \texttt{Herwig7-H7}.
A comparison of the sphericity and $B^+_u$ multiplicity observables is shown in Fig.~\ref{fig:resdistComp} together with their tuning uncertainty as discussed in Section~\ref{sec:uncertainties}. It is noticeable that this uncertainty is systematically larger for the \texttt{Herwig7-H7} tune. This behavior is, at least in part, expected as the larger number of parameters leads to larger uncertainties, even if those are individually constrained to the same degree. The predictions from the models are in agreement within their uncertainties for the sphericity observable, while the central values of the \texttt{Herwig7-P8} model better reproduce the data.
However, the multiplicity, shown in Fig.~\ref{fig:resdistComp}, is described better by the \texttt{Herwig7-H7} model.
In conclusion, both models have similar performance with some observables being better described by either of them. Overall, the \texttt{Herwig7-P8} model has a marginally better agreement with the data in this exemplary set of observables.

\FIGnine{Distributions of $p$-values for the tuned \texttt{Herwig7-H7} and \texttt{Herwig7-P8} hadronization models. Each observable contributes one $p$-value toward the histogram. The bottom figure shows a subrange $p > 10^{-4}$ of the distribution in a logarithmic  scale.}{fig:resultsprobComp}

\FIGten{Distribution of the sphericity observable from DELPHI~\protect\cite{DELPHI:1996sen} on the left and the mean $B^+_u$ multiplicity~\protect\cite{ParticleDataGroup:2008zun} on the right for the data and the tuned MCEG samples for the \texttt{Herwig7-H7} and \texttt{Herwig7-P8} hadronization models. See the caption on Fig.~\ref{fig:restunedobs} for further details. }{fig:resdistComp}

\section{Studies of correlations}
\label{sec:correlation}
For the tuning presented in Section~\ref{sec:results} the uncertainties were assumed to be uncorrelated. In general however, it is expected that uncertainties, especially systematic uncertainties, are in fact correlated to some degree. In order to evaluate the impact of such correlations, additional tunes were performed. The main difference compared to the tunes in Section~\ref{sec:results} is that the covariance matrix $M$ in Eq.~\eqref{eqn:eftfitterl} is no longer purely diagonal. The covariance matrix is now constructed to be blockwise diagonal with a block for each observable distribution. 
For the off-diagonal entries in each block, the product $r\cdot\sigma_i\sigma_j$ was inserted, where $\sigma_{i,j}$ are the systematic uncertainties of data points $i$ and $j$, and $r$ quantifies the amount of correlation between both in terms of a linear correlation coefficient. 
Due to the lack of information regarding the correlation of systematic uncertainties from the analyses, a pragmatic approach was chosen by scanning through the values of the correlation factor $r = [0.0,0.4,0.6,0.8,0.9]$.

While these values are merely suggestive, they represent scenarios with mild, medium and strong correlations. As an example, Fig.~\ref{fig:corfull} shows the two-dimensional posterior probability distribution for the two parameters \texttt{ClMax} and \texttt{ClPow} for different assumptions about $r$. The smallest areas containing 68\% of the marginalized posterior shrink with increasing correlation factor $r$. For large correlation coefficients, such as $r = 0.9$, this area splits into two parts indicating the presence of a second mode within the posterior distribution. 

Introducing and increasing the correlation of uncertainties predominantly results in a narrowing of the posterior phase space, i.e.\ reduced uncertainties of the global modes of the MCEG parameters.
The global mode for the MCEG parameters remains stable with only small deviations which are within the parameter uncertainties.
This behavior of the global mode can be seen in Tab.~\ref{table:corrMode} where the global mode and the standard deviation of the parameters are listed for $r = 0.0$ and $r = 0.9$. Hence, the results of the tune can be regarded as stable, while the estimation of uncertainties of the tuned parameter depends on the choice of correlation factor. 

\FIGeleven{Marginalized two-dimensional distribution of the posterior probability for the parameters \texttt{ClMax} and \texttt{ClPow} for the \texttt{Herwig7-H7} tune. The underlying distribution represents the contours of the posterior for uncorrelated uncertainties with the green, yellow and red contours representing the smallest intervals containing $68$\%, $95$\% and $99$\%, respectively. The overlayed colored contours represent the smallest interval containing $68$\% of the posterior for different configurations of correlation coefficients.}{fig:corfull}

\TABcorrMode

\section{Studies on the effects of weighting observables}

Traditionally, MC tuning relies on the use of weights to stabilize the tuning process and/or increase the importance of certain observables~\cite{Bellm:2019owc}.
This procedure, however, compromises the statistical interpretation of the resulting uncertainties and potentially biases the obtained results of the tune. To test the effect, non-unity weights were introduced in the study and 
the results  were compared to the tune discussed above with all weights equal to unity.
The weights were introduced by adding a vector of coefficients $\vec{w}_{i=1\dots N_{bins}}$ into the Eq.~\eqref{eqn:eftfitterl} and dividing the likelihood by the sum of those weights. All weights for the bins from the same distributions were set to be equal.

In order to investigate the impact of weighting on the tune, the tuning process was repeated for two different weighting schemes $w_1$ and $w_2$ with the weight values for these schemes given in Tabs.~\ref{table:longlistone},~\ref{table:longlistthree} and  ~\ref{table:longlisttwo} in Appendix~\ref{sec:apx_observables}.
The weighting scheme $w_1$ applies higher weights for multiplicities while leaving the event shape variables mostly unchanged.
The weighting scheme $w_2$ sets the weights of the multiplicities to zero and in contrast, increases the weights of the event shape variables and the weights of the mean charged multiplicities observable.

While changing from the unweighted tune to the weighting schemes $w_1$ and $w_2$, most of the posteriors retain their shape with only very minor changes. Most notably the positions of the global modes were shifted by the weighting procedure, as seen in Tabs.~\ref{table:weightHerBig} and \ref{table:weightPytBig}. The mode values of these weighted tunes were also used for the MCEG to compare to data similarly to Section~\ref{sec:results}.
The $p$-values were calculated for the \texttt{Herwig7-H7} and \texttt{Herwig7-P8} tunes for both weighting schemes.

\TABweightHerBig

\TABweightPytBig

The \texttt{Herwig7-H7} tune seems to benefit from the first weighting scheme as the mean of $p$-values increases from $0.133$ to $0.151$ while the second scheme decreases the overall agreement to data. In contrast, the \texttt{Herwig7-P8} benefits from both weighting schemes, although, only yielding a small additional improvement from $0.143$ to $0.147$ when compared to the unweighted tunes from Section~\ref{sec:results}. The posterior distributions for the \texttt{Herwig7-H7} and \texttt{Herwig7-P8} tune are shown in Fig.~\ref{fig:resHerWei} and Fig.~\ref{fig:resPytWei} respectively in Appendix~\ref{sec:apx_corr}.

\section{Conclusions}

A MCEG tuning procedure based on proper statistical grounds using a Bayesian approach is presented.
As an application example, the procedure was used to tune the \texttt{Herwig7} MCEG with two different hadronization models.
The data used for those studies were collected by the LEP experiments and include event-shape and jet rate distributions, charged hadron momentum spectra and multiplicities from the process $e^{+}e^{-}\rightarrow (Z/\gamma)^*\rightarrow {\rm hadrons}$. The \texttt{Rivet} framework was used to generate the analysis code.
The global mode values of the posterior distribution were chosen as the optimal parameter set. In addition, appropriate uncertainty measures for the individual parameters are provided, e.g.\ from the smallest $68$\% credibility intervals, as well as uncertainty propagation of these parameters. The impact of different correlation assumptions and data-weighting schemes on the final results was investigated.

Several conclusions can be drawn from these observations. First it was found that the Bayesian approach to tuning MCEG works successfully. In particular, sets of  optimized parameters including appropriate uncertainty estimates were obtained. These estimates can then be used to propagate the uncertainties on the optimal parameter set to the space of observables. As a result, the tuning of MCEG improves the agreement between the data and the MCEG predictions, and it provides an estimate for the uncertainties of prediction from the tuning process itself. 

Second, it was shown that correlations between measurements can have an impact on optimization results, especially when uncertainties are estimated. Therefore it is recommended to the experimental collaborations if possible to derive and publish those correlations, and suggest that such correlations be carefully considered in further tuning studies.
Third, it was found that for the particular data sets, the MCEG models examined, and the tuning parameters considered, the \texttt{Herwig7} MCEG using the Lund string hadronization model describes the data slightly better than the standard cluster hadronization model.

\section*{Acknowledgments} 
\label{sec:acknowledgements}

This work was supported by the German Science Foundation DFG through the Collaborative Research Center SFB1491 and project KR 4060/7-1

The authors are grateful to the authors and contributors of the software packages which made this analysis possible. The authors are grateful to Andrzej Siodmok for valuable discussions of this manuscript and the presented analysis.

\FloatBarrier

\appendix
\allowdisplaybreaks
\section{Software used in the analysis}
\label{sec:software}
The list of the used software is given below.
The statistical analysis was performed using the 
\texttt{BAT.jl}~\cite{Schulz:2021BAT} and
\texttt{EFTFitter.jl}~\cite{EFTfitter2016} packages of the 
\texttt{JULIA} language. 
The fitting was performed using the \texttt{LsqFit.jl} package for the \texttt{JULIA} language.
The generation of the MC event samples and their processing was done using 
\texttt{Herwig7.2.2}~\cite{Bellm:2015jjp},
\texttt{MadGraph5}~\cite{Alwall:2011uj},
\texttt{OpenLoops}~\cite{Cascioli:2011va},
\texttt{Pythia8}~\cite{Bierlich:2022pfr},
\texttt{TheP8I}~\cite{TheP8I},
\texttt{Rivet}~\cite{Bierlich:2019rhm},
\texttt{ROOT 6.22}~\cite{Antcheva:2011zz}
and \texttt{HepMC3} packages packed 
into \texttt{singularity} containers based on Fedora Linux distribution.

\FloatBarrier
\newpage
\section{Observables used in the analyses}
\label{sec:apx_observables}
\TABlonglistone{Lists of \texttt{Rivet} analyses and bins used in the tuning including their descriptions imported from \texttt{Rivet} as well as two different weighting schemes used for different tunes, part I~\protect\cite{ALEPH:1996oqp,ALEPH:2001pfo,JADE:1999zar}.}
\FloatBarrier

\TABlonglisttwonew{Lists of \texttt{Rivet} analyses and bins used in the tuning including their descriptions imported from \texttt{Rivet} as well as two different weighting schemes used for different tunes, part II~\protect\cite{ParticleDataGroup:2008zun}.}

\TABlonglistthree{Lists of \texttt{Rivet} analyses and bins used in the tuning including their descriptions imported from \texttt{Rivet} as well as two different weighting schemes used for different tunes, part III~\protect\cite{DELPHI:1996sen}.}

\FloatBarrier

\section{Studies of the weights impact}
\label{sec:apx_corr}
\FIGtwelve{Distribution of $p$-values for tuned and nominal MC samples for the \texttt{Herwig7-H7} tune with weighting scheme one (two top distributions) and two (two bottom distributions). Each observable contributes one $p$-value toward the histogram. The bottom figures show a subrange $p > 10^{-4}$ of the distributions  in a logarithmic  scale.}{fig:resHerprobweight}
\FIGthirteen{Distribution of $p$-values for tuned and nominal MC samples for the \texttt{Herwig7-P8} tune with weighting scheme one (two top distributions) and two (two bottom distributions). Each observable contributes one $p$-value toward the histogram. The bottom figures show a subrange $p > 10^{-4}$ of the distributions  in a logarithmic  scale.}{fig:resPytprobweight}
\FloatBarrier
\FIGfourteen{One and two dimensional marginalized posterior distributions for the \texttt{Herwig7-H7} tune using weighting scheme one (top) and two (bottom). See the caption on Fig.\ref{fig:resultsHer} for a detailed explanation.}{fig:resHerWei}
\FIGfifteen{One and two dimensional marginalized posterior distributions for the \texttt{Herwig7-P8} tune using weighting scheme one (top) and two (bottom). See the caption on Fig.\ref{fig:resultsHer} for a detailed  explanation. }{fig:resPytWei}

\FloatBarrier

\section{Studies on the goodness of fit}
\label{sec:apx_ipol}


The test samples, described in Sec.~\ref{sec:Approximation} were used to study the approximation model. 
The performance of the approximation model was evaluated using the residuals between the 
approximated values and the analysis and test samples.
The corresponding pulls for those residuals in the analysis samples were calculated for each fitted bin as
\begin{equation}
    p_{\rm analysis} = \frac{f(\vec{\lambda})-D_{MC}}{\sigma_r} = \frac{f(\vec{\lambda})-D_{MC}}{\sqrt{\sigma_{D_{MC}}^2-\sigma_{f}^2}}
    \label{eqn:pull}
\end{equation}
with the MCEG values $D_{MC}$, their uncertainties $\sigma_{D_{MC}}$, the value of the approximation $f(\vec{\lambda})$ and the uncertainty of the approximation $\sigma_{f}^2$ which was calculated by propagating the fit uncertainties on the coefficients~\cite{BlobelLohrmann:2012}.
The calculations of pulls with the test samples take into account that the
uncertainties of $D_{MC}$ and $f$ are uncorrelated, hereby
\begin{equation}
    p_{\rm test} = \frac{f(\vec{\lambda})-D_{MC}}{\sigma_r} = \frac{f(\vec{\lambda})-D_{MC}}{\sqrt{\sigma_{D_{MC}}^2+\sigma_{f}^2}}
    \label{eqn:pulltwo}
\end{equation}
In both cases, in Eq.~\ref{eqn:pull} and Eq.~\ref{eqn:pulltwo} it is possible for the uncertainty of the fit to become larger than the uncertainty of the MCEG data points as the error propagation relies on a linear approximation of the cubic model. Hence,
data points with $\sigma_{f}^2 > \sigma_{D_{MC}}^2$ were omitted from the pull distributions and the behaviour of the approximation model for those points was studied on a case-by-case basis.
The $p_{\rm test}$ and $p_{\rm analysis}$ follow normal distributions, although the $p_{\rm analysis}$ distribution shows larger values toward the tails. 
In addition, normal distributions were fitted to  $p_{\rm analysis}$ and $p_{\rm test}$, which resulted in mean values of $-0.008$ and $-0.003$, respectively. The fitted standard deviations are $1.66$  and $1.18$.
The higher standard deviation  and the larger tails of the $p_{\rm analysis}$ distribution  indicate that the fitted approximation models do not always give a perfect description of the MCEG data, leading to larger average residuals than expected from statistical fluctuations alone. 
While it is beneficial for further tuning efforts to refine the description of the MCEG data, the approximation model suffices for the study's emphasis on presenting a new tuning method.

Apart from the checks described in Sec.~\ref{sec:Approximation},  additional studies were performed to ensure that the approximation procedure works as expected.
The $\chi^2_\text{red}=\chi^2/n_\text{dof}$ values for each bin were used to evaluate the performance of the approximation procedure. 
The distribution of $\chi^2_\text{red}$ for all approximated observables with the \texttt{Herwig7} MCEG is given in Fig.~\ref{fig:chisqfit}.
The distribution peaks at the $\chi^2_\text{red}\approx 1$, with most of the entries having $\chi^2_\text{red} < 15$. 
The tail of the distribution contains entries from the presumably poorly approximated bins, however, the number of those entries is quite small, which is a desirable property for the approximation procedure.

\FIGsixteen{Distribution of $\chi^2_\text{red}$ for all the approximated observables and bins. Each bin of each observable contributes one value to the distribution.}{fig:chisqfit}

\FIGseventeen{Calculated $p$-values for the $\chi^2$ tests on the test samples evaluated for the \texttt{Herwig7} MCEG.}{fig:gridpval}

The $p$-values for the $\chi^2$ tests for the test samples for the \texttt{Herwig7} MCEG are calculated and are shown in Fig.~\ref{fig:gridpval}. 
Both the $\chi^2_\text{red}$ and $p$-values indicate potential issues with the approximation for certain bins, however, they do not provide a quantitative goodness of fit result.
Hence, bins of observables with low $p$-values and large $\chi^2_\text{red}$ values were further inspected manually using the test samples, e.g.\ considering distributions similar to those shown in Fig.~\ref{fig:gridfit}.

For most of those bins, the approximated values are compatible with the MC values if the variances introduced by the choice of the default parameters are taken into account.
However, due to the limited number of parameters used in the approximation it is not possible to reproduce the MC samples exactly.